\documentclass[a4paper,aps,reprint,pra,11pt, onecolumn, superscriptaddress,nofootinbib, showpacs, notitlepage, accepted=2024-03-25]{quantumarticle}

\pdfoutput=1

\usepackage[pdftex]{graphicx}
\usepackage[dvipsnames]{xcolor}
\usepackage{ulem}
\usepackage{color}
\usepackage{amsmath}
\usepackage{amssymb}
\usepackage{dcolumn}
\usepackage{bbold}
\usepackage{braket}
\usepackage{lipsum} 
\usepackage{caption}
\usepackage{CJKutf8}
\usepackage{subcaption}
\usepackage{amsthm}
\usepackage{setspace}
\usepackage{hyperref}
\usepackage[T1]{fontenc}

\usepackage{url}

\begin{document}

\title{Relational superposition measurements with a material quantum ruler}

\begin{CJK*}{UTF8}{bsmi}

\author{Hui Wang (王惠)}
\email{huiwangph@gmail.com}
\affiliation{Department of Physics and Astronomy, Dartmouth College, Hanover, New Hampshire 03755, USA}
\affiliation{Theoretical Quantum Physics Laboratory, Cluster for Pioneering Research, RIKEN, Wako-shi, Saitama 351-0198, Japan}
\orcid{0000-0003-1012-1124}

\author{Flaminia Giacomini}
%\email{fgiacomini@phys.ethz.ch}
\affiliation{Perimeter Institute for Theoretical Physics, 31 Caroline St. N, Waterloo, Ontario, N2L 2Y5, Canada}
\affiliation{Institute for Theoretical Physics, ETH Z{\"u}rich, Wolfgang-Pauli-Str. 27, Z{\"u}rich, Switzerland}
\orcid{0000-0002-2475-2296}

\author{Franco Nori (野理)}
%\email{fnori@riken.jp}
\affiliation{Theoretical Quantum Physics Laboratory, Cluster for Pioneering Research, RIKEN, Wako-shi, Saitama 351-0198, Japan}
\affiliation{Quantum Computing Center, RIKEN, Wako-shi, Saitama 351-0198, Japan} %
\affiliation{Department of Physics, University of Michigan, Ann Arbor, Michigan 48109-1040, USA}
\orcid{0000-0003-3682-7432}

\author{Miles P. Blencowe}
%\email{Miles.P.Blencowe@dartmouth.edu}
\affiliation{Department of Physics and Astronomy, Dartmouth College, Hanover, New Hampshire 03755, USA}
\orcid{0000-0003-0051-0492}

\maketitle
\end{CJK*}

\begin{abstract}
In physics, it is crucial to identify operational measurement procedures to give physical meaning to abstract quantities. There has been significant effort to define time operationally using quantum systems, but the same has not been achieved for space. Developing an operational procedure to obtain information about the location of a quantum system is particularly important for a theory combining general relativity and quantum theory, which cannot rest on the classical notion of spacetime. 

Here, we take a first step towards this goal, and introduce a model to describe an extended material quantum system working as a position measurement device. Such a ``quantum ruler'' is composed of $N$ harmonically interacting dipoles and serves as a (quantum) reference system for the position of another quantum system. 

We show that we can define a quantum measurement procedure corresponding to the ``superposition of positions'', and that by performing this measurement we can distinguish when the quantum system is in a coherent or incoherent superposition in the position basis. The model is fully relational, because the only meaningful variables are the relative positions between the ruler and the system, and the measurement is expressed in terms of an interaction between the measurement device and the measured system. 
\end{abstract}
\maketitle

\tableofcontents

\section{Introduction}

In physics, operational measurement procedures give physical meaning to observable quantities in terms of laboratory operations. Such procedures require concrete models of measurement devices, and a realistic description of interactions between the device and the measured system. The definition of such procedures is important not only for practical purposes, but also from a fundamental perspective: in experimentally unexplored regimes of physics the correspondence between abstract observables and physically meaningful quantities can be ambiguous. For instance, in special-relativistic quantum physics this is the case for the relativistic spin operator~\cite{Peres:2002ip, bauke2014relativistic}. More strikingly, at the interface between quantum theory and gravity basic notions such as time, space, and causality cannot be kept unchanged. Hence, it is crucial to develop methods to characterise physically meaningful quantities via procedures that can then be employed in more general scenarios than those currently tested experimentally.

An operational definition of time can be obtained using quantum clocks. In quantum information, quantum clocks have been studied, e.g., in relation to thermodynamics~\cite{erker2017autonomous} and to the possibility of measuring time more accurately than with classical clocks~\cite{woods2018quantum}.  In gravity, quantum clocks constitute a promising tool to investigate the properties of physics at the interface between quantum theory and gravity~\cite{Tino:2007az, zych_interferometric, Giulini:2011xf, zych, Roura:2018cfg}.

Despite the attention that operational procedures to measure time have received, not much is known about analogous procedures to measure spatial positions. To overcome idealised position measurements, in which the reading in the laboratory frame corresponds to our standard notion of distance, we need a concrete model of a ``quantum ruler'', namely a quantum system that provides us with information on the position of another quantum system, i.e. the measured system. 

\begin{figure}[t]
\begin{center}
\includegraphics[width=3.5in]{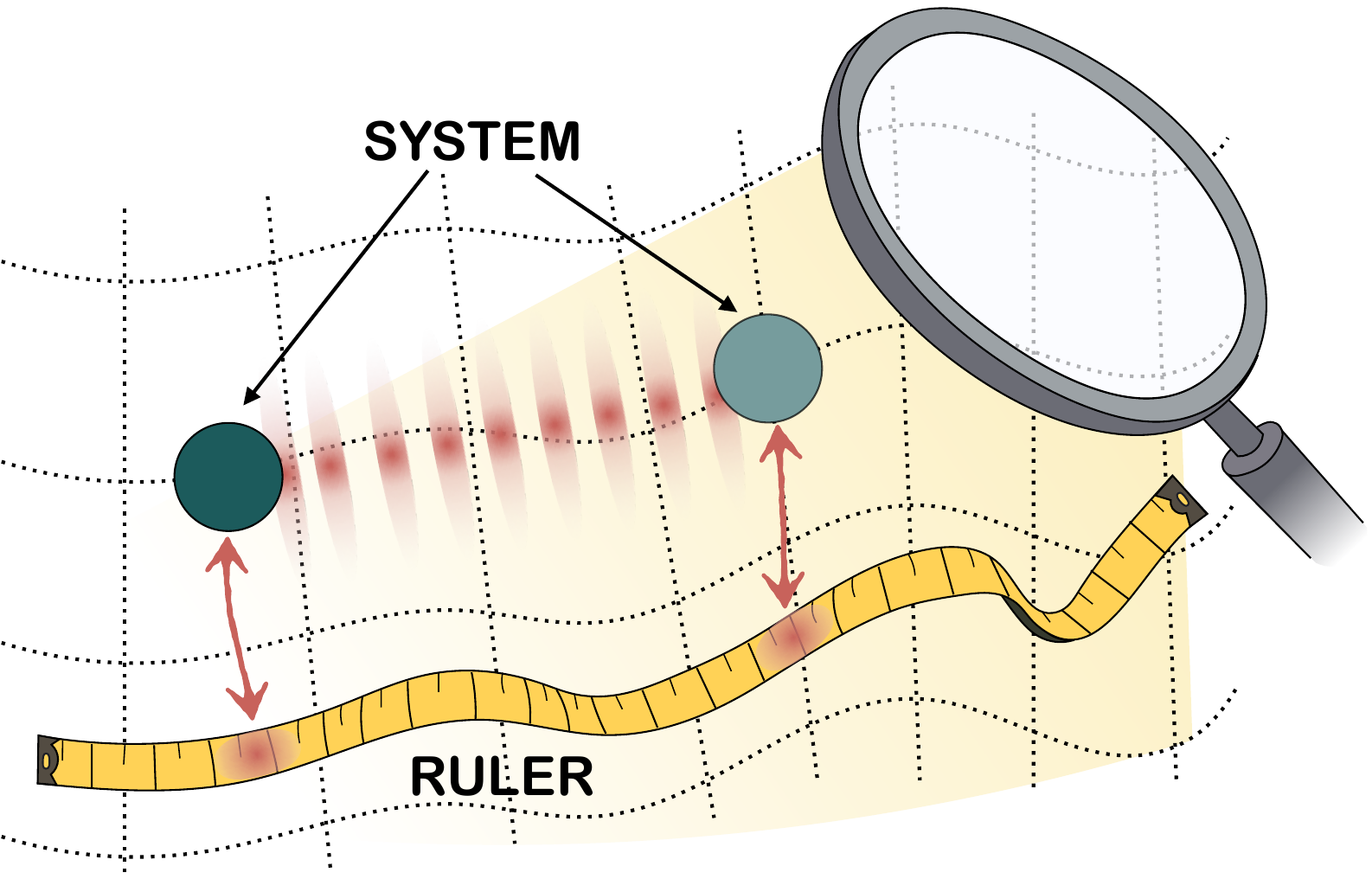} \caption{\label{fig:cartoon} Intuitive illustration of the functioning of the quantum ruler. We introduce a relational position measurement scheme for systems in a spatial quantum superposition, which does not depend on any abstract or absolute quantity and is solely expressed in terms of relations between two physical systems. Central to the idea is to develop a concrete model of a quantum ruler which interacts (red arrows) with a quantum system initially prepared in a quantum superposition state of two different locations. The ruler is distorted (red spots) as a result of the interaction with the system. We show that after the measurement, which involves both the ruler and the quantum system, the coherence of the quantum superposition  (the interference pattern) can be preserved.} 
\end{center}
\end{figure}

Here, we introduce such a quantum ruler (Fig. \ref{fig:cartoon}). A quantum ruler has been previously considered in Refs.\,\cite{Knight:1996sxc, ralph2002coherent} in a different context and with different goals to the one we have here. To make our quantum formulation amenable to a general relativistic description, the ruler should be an extended quantum system, which serves as the reference system for the measurement of positions. Considering the ruler as a physical system allows us to define local observables that are also background independent. In the quantum gravity literature, the expectation that a theory of quantum gravity should be background independent has been related to  the necessity of considering extended material reference frames~\cite{dewitt1967quantum, Kuchar:1990vy, Brown:1994py, Brown:1995fj}, such as an elastic medium. Physically, this means that the observables (intended in a broad sense as measurable quantities) should be relational. Hence, they should be expressed in terms of the interaction between two physical systems and not rely on any background or absolute structure~\cite{butterfield2001spacetime, Rovelli:2004tv, rovelli_relational, anderson2017problem}.

Our quantum ruler is composed of $N$ identical electric dipoles, which are coupled with a harmonic potential to their neighbouring ones, forming a one-dimensional ``mass-spring chain''. The measured system is an ion, which can be initially prepared in a localised state, in a mixed state of two positions, or in a pure quantum superposition state in the position basis. 

While the considered setup is to be viewed as a simplified model of a more realistic, three-dimensional extended material ruler, one-dimensional trapped ion \cite{cirac1995,chen2023} and dipole atom chain \cite{vetsch2010} systems with phononic modes have been considered in the lab for quantum information processing applications. Our motivation for considering a dipole rather than an ion chain model, is to have a more local interaction between the ion system and ruler.

We show that the ion system-dipole atom ruler can be approximated using the Hamiltonian of a multi-mode optomechanical system with $N-1$ mechanical oscillators, allowing exact solutions for the resulting ion-ruler quantum dynamics \cite{bruschi2019time}. When considering a large number of dipoles ($N\gg1$), the ruler behaves as an oscillator ``bath'' environment for the ion system. Thus, if we restrict ourselves to measurements of the ion system only, 
%by discarding the bath environment (the ruler)
the ion's state will decohere \cite{zurek1991} in the position basis.

Our goal is to construct a measurement scheme to obtain information about the position of the ion without losing the coherence of the quantum state. This measurement is more general than the usual position measurements, in that it measures the system in a ``quantum superposition of positions''. Crucially, this procedure is expressed solely in terms of relative quantities, and the results do not depend on the state nor on the dynamics of the centre of mass degree of freedom of the ruler.

The generalisation of such a measurement encounters nontrivial challenges, which are due to the complexity of the quantum ruler as a one-dimensional many-body system. In particular, we want to reduce all possible decoherence effects on the state of the ion. We show that an appropriate choice of parameters of the ruler and measurement procedures ensures that measurements via the quantum ruler can distinguish a pure quantum superposition state of the ion from an incoherent mixture. We also comment on how future work could develop and enrich the measurement scheme we introduce here.

The paper is organised as follows. In Section~\ref{sec:ToyModel} we introduce a relational toy-model of $N$ non-relativistic particles, which captures the main conceptual features of the ruler, but is technically much simpler. In Section~\ref{sec:ToyMeas} we develop a method to construct relational position measurements. In Section~\ref{sec:quantumruler} we introduce the quantum ruler, and in Section~\ref{sec:measurement} we illustrate the measurement procedure involving the quantum ruler and the ion.

\section{A relational toy model with classical particles}
\label{sec:ToyModel}
 
We consider $N$ non-relativistic particles with mass $m_\alpha$ and coordinates $q_\alpha$, with $\alpha=1,\cdots, N$. The dynamics is governed by the Lagrangian
\begin{equation} \label{eq:LagNpart}
	\mathcal{L} = \sum_{\alpha=1}^N \frac{m_\alpha}{2} \dot{q}_\alpha^2 - \sum_{\alpha=1}^{N-1} V_\alpha ( q_{\alpha+1} - q_\alpha)- \frac{1}{2M}\left( \sum_\alpha m_\alpha \dot{q}_\alpha \right)^2,
\end{equation}
where $M = \sum_\alpha m_\alpha$ is the total mass, the first term is the kinetic energy of the $N$ particles, and the second term is an interacting potential between two neighbouring particles, which scales with the relative distance between the particles ($q_{\alpha+1} - q_\alpha$). The last term rescales the total energy by subtracting the energy of the centre of mass. Equivalently, this term makes the system fully translationally invariant, so that the relative velocities between two particles are the only meaningful quantities. The kinetic term can be cast as~\cite{Lynden-Bell:1995cmj}
\begin{equation}
    T = \sum_{\alpha=1}^N \frac{m_\alpha}{2} \dot{q}_\alpha^2 - \frac{1}{2M}\left( \sum_\alpha m_\alpha \dot{q}_\alpha \right)^2 = \sum_{\alpha=1}^N \frac{m_\alpha}{2}\left(\dot{q}_\alpha - \dot{q}_{CM}\right)^2,
\end{equation}
where $q_{CM}= \sum_\alpha \frac{m_\alpha}{M} q_\alpha$. We now make a coordinate transformation to the centre of mass coordinates and the relative coordinates of the $N$ particles to the centre of mass: 
\begin{equation} \label{eq:coordCMrel}
    x_{CM} = q_{CM}= \sum_\alpha \frac{m_\alpha}{M} q_\alpha, \qquad x_\alpha = q_\alpha - q_{CM},
\end{equation}
with $\alpha=1, \cdots, N$. Notice that in this step we have introduced a redundant coordinate, which we will eliminate later using the identity $\sum_{\alpha=1}^N m_\alpha x_\alpha = 0$. The Lagrangian can then be expressed in this set of coordinates as
\begin{equation} \label{eq:LagN+1}
	\mathcal{L} = \sum_{\alpha=1}^N \frac{m_\alpha}{2} \dot{x}_\alpha^2  - \sum_{\alpha=1}^{N-1} V_\alpha ( x_{\alpha+1} - x_\alpha).
\end{equation}
The canonical momenta are
\begin{equation}
    \pi_\alpha = \frac{\partial \mathcal{L}}{\partial \dot{x}_\alpha}= m_\alpha \dot{x}_\alpha, \qquad \pi_{CM} = \frac{\partial \mathcal{L}}{\partial \dot{x}_{CM}} = 0.
\end{equation}
The Hamiltonian corresponding to the Lagrangian of the system is then
\begin{equation} \label{eq:HC1}
     H = \sum_\alpha \frac{\pi_\alpha^2}{2 m_\alpha} + \sum_{\alpha=1}^{N-1} V_\alpha ( x_{\alpha+1} - x_\alpha) + \mu \mathcal{C},
\end{equation}
where $\mu$ is a Lagrange multiplier and $ \mathcal{C}= \pi_{CM}$ is a constraint coming from the equations of motion, i.e.\,$ \mathcal{C} \approx 0$\footnote{The symbol $\approx$ denotes a ``weak equality'', namely an equality that holds on the constraint surface.}. This constraint is trivially satisfied, as the Hamiltonian does not depend on it. Finally, we need to impose the identity $\sum_{\alpha=1}^N m_\alpha x_\alpha = 0$. This can be easily done by eliminating one of the particles from the description. For instance, we can choose to remove particle 1 and write $x_1 = -\sum_{\alpha=2}^N \frac{m_\alpha}{m_1} x_\alpha$. Notice that this condition should not be treated as a dynamical constraint, because it is an artefact of our coordinate transformation. Hence, it holds in general, and not only on the space of solutions of the equations of motion, and can be simultaneously implemented with $\mathcal{C} \approx 0$.

\section{Position measurements with a material reference frame}
\label{sec:ToyMeas}

It is generally believed that diffeomorphism invariance is not compatible with the possibility of defining meaningful local measurements. However, this problem is solved when one abandons the abstract notion of a coordinate system and considers material reference frames, namely reference frames associated to physical (matter) systems. In this case, a measurable quantity $O$ is not evaluated at an abstract point of a manifold $\mathcal{M}$, i.e.\,$O(x)$, with $x\in \mathcal{M}$, but should be considered as an ``event'' arising from the interaction between two physical systems. As a consequence, when a diffeomorphism transformation is performed, both the material reference frame and the observed system are transformed, and diffeomorphism invariance is preserved.

In quantum theory, it is possible to construct local observables as quantum operators that are invariant under a gauge (diffeomorphism) transformation. In quantum gravity, relational observables have been defined as Dirac observables by employing an extended, material coordinate system~\cite{dewitt1967quantum, Dittrich:2004cb, Tambornino:2011vg, Hoehn:2019fsy, Giesel:2020xnb, Bojowald:2021uqo, Baldazzi:2021fye}. 

Usually, a position measurement is defined by choosing an abstract coordinate system labelled by $x$ and measuring an observable $\hat{O}_S$ acting on a system $S$. If $\hat{O}_S$ is the position operator, then the measurement returns outputs some classical value $x^*$, which is the position of the system on the abstract coordinate system. This procedure is represented as a set of projective operators: for each possible value $x^*$ of the position measurement, there is a corresponding projective measurement operator
\begin{equation}
    \hat{O}_S(x^*) \rightarrow \hat{\Pi}_{x^*} = \ket{x^*}_S\bra{x^*}.
\end{equation}
If we change the coordinate system, this local measurement is not diffeormorphism invariant, because each value $x^*$ is transformed by the diffeormorphism transformation. In a relational picture, all positions are instead the relative distance between two physical systems, one of which, $r$, serves as the reference system. In our case, $r$ is a ruler, described as an extended quantum system on a lattice, where the $N$ lattice sites are labelled as $i=1, \cdots, N$. The ruler $r$ interacts with the system $S$ via the unitary operator $U_{Sr}=e^{-i\epsilon \hat{G}_{Sr}}$, where $\hat{G}_{Sr}$ denotes the system-ruler interaction Hamiltonian, and $\epsilon$ represents the interaction strength. The effect of this interaction is to entangle the system of the ruler and system. With this procedure, the position measurement becomes a joint measurement of the system and the ruler, denoted as $\hat{O}_{Sr}$:
\begin{equation} \label{eq:ObsRel}
  \hat{O}_{Sr}(s_i) \rightarrow \hat{\Pi}_i \otimes \ket{s_i}_r\bra{s_i},
\end{equation}
where $i$ labels the site of the ruler, and not an abstract coordinate system. Notice that the ruler is constructed to be invariant under global translations, so that it does not matter\footnote{A subtlety is to make sure that the measurement is not affected by boundary size effects, meaning that the system $S$ should be distant from the edges of the ruler. We discuss this point in Section~\ref{sec:quantumruler}.} where the quantum system is relative to the ruler.

Our goal here is to construct a measurement corresponding to the ``quantum superposition of positions''. A minimal requirement that we impose is that such a measurement is a Positive Operator Valued Measure (POVM). If we were to use an abstract coordinate system, the most natural choice for a POVM would be to divide the spatial extension of the laboratory into $N$ slots of length $\Delta m$. Calling $R =\Delta m$ the resolution of the measurement apparatus, it is then possible to check in which slot the system is found\footnote{A similar strategy was used in Refs.~\cite{kofler2007classical, dakic2016classical}, for finite-dimensional systems, to define the classical limit of a quantum theory.}. However, this procedure is suitable to measure the system in a well-defined position, but cannot measure the system in a quantum superposition of positions, because no set of POVM corresponding to the quantum superposition of positions can be defined on the single system $S$. However, this can be achieved by including the ruler in the picture and defining a measurement acting on both the system $S$ and the ruler $r$ as in Eq.\,\eqref{eq:ObsRel}. This allows us at once to i) consider more general position measurements corresponding to projective measurements, and ii) establish a relational, local picture in which position measurements can be performed.

We here outline an intuitive description of this procedure by providing a simplified model of the ruler. Key to this procedure is the existence of a suitable interaction between the ruler and the system which gives rise to an entangled state of $S$ and $r$. We give the explicit form of the interaction in Section~\ref{sec:quantumruler} and of the measurement in Section~\ref{sec:measurement} for the actual quantum ruler.

In this idealised description of the quantum ruler, each site $m$ is a two-level system, where the state $\ket{0}_{r_m}$ corresponds to the ruler not being distorted and the state $\ket{1}_{r_m}$ corresponds to the ruler being distorted by the interaction with the quantum system $S$. A general state of the ruler is then
\begin{equation}
	\ket{\Psi}_r = \sum_{s_1, s_2, \cdots, s_N=0,1} c_{s_1 s_2 \cdots s_N} \ket{s_1}_{r_1} \ket{s_2}_{r_2} \cdots \ket{s_N}_{r_N}.
\end{equation}
In an ideal scenario where the system is localized near ruler site $j$, a distortion occurs at the ruler site $j$. As we explain in Section IV, this is not realistic for the actual quantum ruler, which responds to the interaction with the system in a non-local way. We define the state in which one site is distorted as $\ket{0}_{r_1} \ket{0}_{r_2} \cdots \ket{1}_{r_j} \cdots \ket{0}_{r_N} = \ket{j=1}_r$. The new single-position measurement is an operator on the Hilbert space of the ruler and of the quantum system, namely
 \begin{equation} \label{eq:POVM}
 	M_{j} = \ket{\psi_{j}}_S \bra{\psi_{j}} \otimes \ket{j=1}_r \bra{j=1},
 \end{equation}
 where $\ket{\psi_{j}}_S$ is a normalised state on the Hilbert space of the quantum system centred at the $j$th site and roughly constant in the corresponding slot, and having typical width $\sigma$ which is at least of the order of the measurement apparatus resolution $R$. For instance, they could be coherent states with very large $\sigma$. Notice that, in the case of an ideal ruler, we do not need the states $\ket{\psi_{j}}_S$ to be orthogonal for different values of $j$, because the measurement acting on the dipoles of the ruler ensures that $M_{j} M_{k} = M_{j} \delta_{j, k}$. However, the states $\ket{\psi_{j}}_S$ must form a basis of the Hilbert space of the system $S$. In addition, we also obtain the completeness condition $\sum_m M_m + \sum_{\bar{m}} M_{\bar{m}} = \mathbb{1}_{\text{lab}}$, where the first term includes the measurements that are physically relevant for us (for instance, those corresponding to the slot where the system is) and $\bar{m}$ labels the complementary set. Here, $\mathbb{1}_{\text{lab}}$ is the identity operator on the Hilbert space of the system and ruler restricted to the laboratory, where the laboratory might comprise the edge of the ruler or other delineated boundaries.

This construction can be extended to measure the system $S$ that is prepared in a quantum superposition of different sites. In this work, we are particularly interested in determining whether the system exists in a superposition involving two distinct sites. Suppose that based on prior knowledge, we learn that the system is restricted to two specific sites, denoted as $j$ and $k$. However, the coherence of the system over the two sites remains uncertain. In such a scenario, the measurement to define the coherence can be described as
\begin{equation} \label{eq:Mpm}
	M_{\pm} = \frac{1}{2} \left[ \ket{\psi_{j}}_S \ket{j=1}_r \pm \ket{\psi_{k}}_S \ket{k=1}_r\right] \left[_S\bra{\psi_{j}}_r \bra{j=1} \pm {}_S\bra{\psi_{k}}_r \bra{k=1}\right].
\end{equation} 
It is then easy to check that the set of measurements $\tilde{M}_{i=1}^N = \Big\lbrace M_+, M_-, \lbrace M_i\rbrace_{i \neq j,k} \Big\rbrace$ satisfies the same relations $\tilde M_{l} \tilde M_{m} = \tilde M_{l} \delta_{l, m}$ and $\sum_m \tilde M_m + \sum_{\bar{m}} \tilde M_{\bar{m}} = \mathbb{1}_{lab} $. In what follows, we consider a regime in which the interaction strength $\epsilon$ between the ruler and the system is small. In this case, the measurement is analogous to a ``weak'' measurement as defined, e.g., in Ref.\,\cite{oreshkov2005weak}.

\section{A material quantum ruler as an extended reference frame}
\label{sec:quantumruler}

\begin{figure}[htbp]
    \centering
    \includegraphics[width=0.8\linewidth]{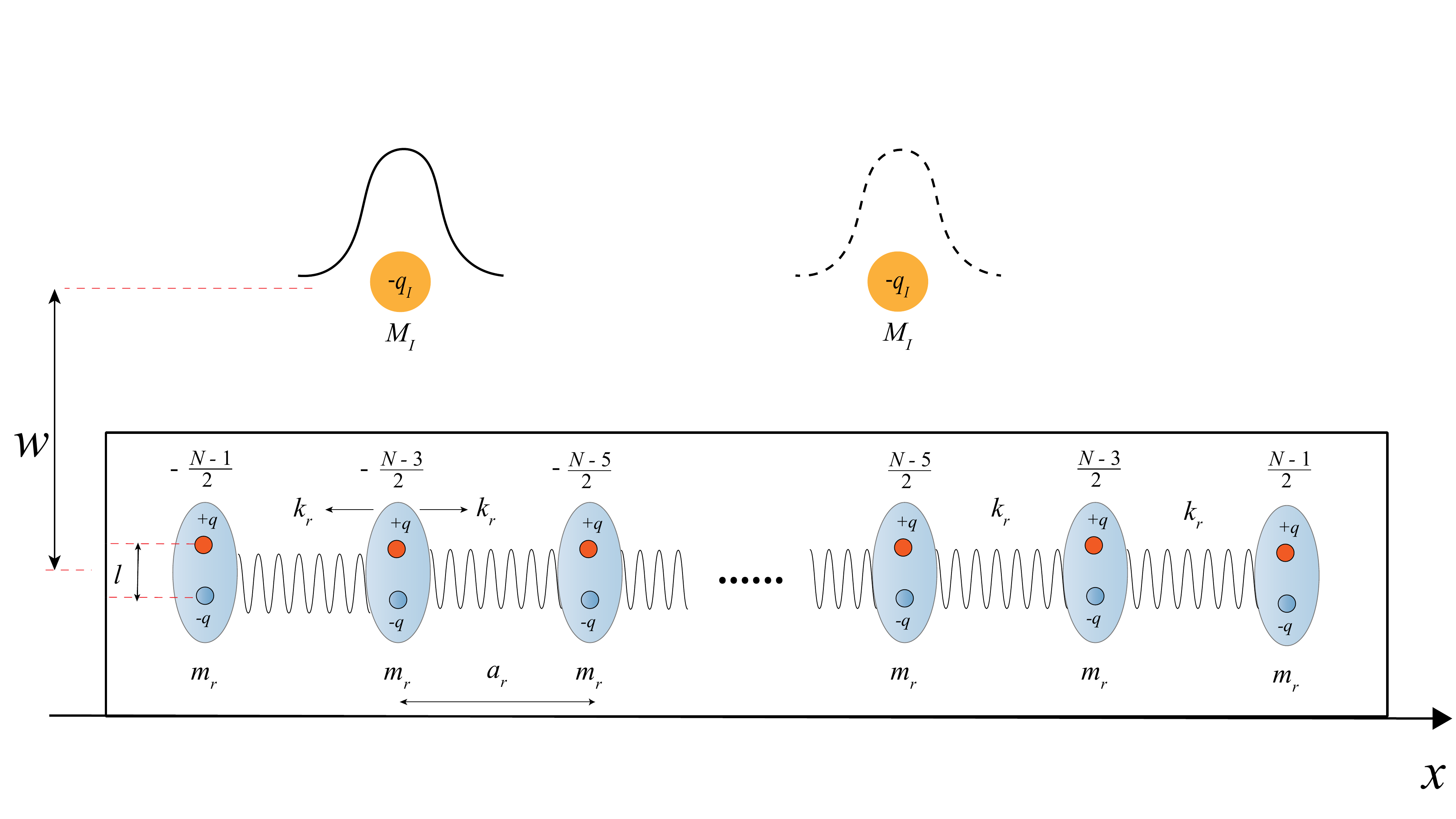}
    \caption{Schema of the one-dimensional quantum ruler model. The ruler is composed of $N$ (odd) electric dipoles that are coupled via harmonic nearest-neighbour interactions (represented by the springs). The dipoles also interact electrostatically with an ion with charge $-q_{I}$, whose state is in a quantum superposition in the position basis, restricted to a parallel axis a distance $w$ from the ruler, and assumed to be far away from the edges of the ruler. The ion induces displacements of the ruler dipoles, from which the position of the ion relative to the ruler can be determined.}
    \label{fig:ionruler}
\end{figure}

%\begin{figure}
%\begin{center}
%\includegraphics[width=6in]{ionruler.pdf} %\caption{\label{fig:ionruler} Schema of the one-dimensional quantum ruler model. The ruler is composed of $N$ (odd) electric dipoles that are coupled via harmonic nearest-neighbour interactions (represented by the springs). The dipoles also interact electrostatically with an ion with charge $-q_{I}$, whose state is in a quantum superposition in the position basis, restricted to a parallel axis a distance $w$ from the ruler, and assumed to be far away from the edges of the ruler. The ion induces displacements of the ruler dipoles, from which the position of the ion relative to the ruler can be determined.} 
%\end{center}
%\end{figure}

We introduce a relational model, illustrated in Fig.~\ref{fig:ionruler}, consisting of a one-dimensional quantum ruler $r$ interacting with a quantum system $I$. The ruler is composed of $N$ identical dipoles of mass $m_r$ (with $N$ odd, so that a dipole is situated at the geometric midpoint of the ruler in classical static equilibrium), which are coupled via a harmonic nearest-neighbour interaction (indicated by the springs connecting the dipoles in the figure) with effective spring constant $k_r$ and classical, static equilibrium separation $a_r$. The quantum system is located at a fixed, vertical distance $w$ from the ruler, but otherwise free to move parallel to the ruler $x$-coordinate axis. We take this quantum system to be an ion (hence the use of the measured system label `$I$') of mass $M_I$, which can, for example, be prepared in a quantum superposition of localised position states. Our goal is to introduce a relational measurement between the ruler and the ion which does not localise the quantum state of the ion, i.e., preserving its superposition properties. Here, by relational we imply that the meaningful coordinates are distances along the $x$-coordinate axis between the physical systems involved, and that the centre of mass does not play a role in our description. In addition, the observables that we measure are relational, in that they provide information about the position of the ion (the measured system) in terms of correlations between its quantum state and the states of the dipoles composing the ruler. In this sense, our model does not require an abstract, background absolute coordinate system for its definition. 

By allowing for the possibility of arbitrarily large (odd) dipole number $N$, our ruler model captures some of the features of actual, macroscopic material extended ruler systems--a primary motivation for the model. The one-dimensional nature of our model is an idealization, however, allowing for exact analytical solutions to the interacting, many-body ion system-ruler dipole quantum dynamics in terms of phonon modes. A price to pay in working with a one-dimensional mass-spring model, however, is that the ruler elastic displacement response to a localised ion state is nonlocal, extending throughout the length of the ruler. A more realistic two or three-dimensional mass-spring lattice model will exhibit more localised distortions opposite the ion localisation, but with the phonon mode dynamics more challenging to analyze.    

The total Hamiltonian operator of the ruler and the ion can be decomposed as $\hat{H}=\hat{H}_r + \hat{H}_{I}  + \hat{V}_{I r}$, where $\hat{H}_r$ is the ruler Hamiltonian:   
\begin{equation} \label{eq:Hrfree}
 \hat{H}_r = \frac{\hat{p}^2_{\mathrm{rCM}}}{2 M_r}+ \sum_{n=-\frac{N-1}{2}}^{\frac{N-1}{2}} \frac{\hat{\pi}_n^2}{2m_r} + \frac{1}{2} k_r \sum_{n=-\frac{N-1}{2}}^{\frac{N-3}{2}} (\hat{\phi}_{n+1} - \hat{\phi}_{n})^2.
\end{equation}
This Hamiltonian has a more transparent physical interpretation in the Lagrangian picture; we summarize our derivation of the Lagrangian in Appendix~\ref{app:Lagrangian}. The first term in the Hamiltonian~\eqref{eq:Hrfree} is the kinetic energy of the ruler centre of mass, with $M_r = Nm_r$ the total ruler mass.  The harmonic  potential energy of the ruler is in terms of nearest-neighbour dipole coordinate differences, with $\hat{\phi}_n = \hat{x}_{r,n} - na_r- \hat{x}_\mathrm{rCM}$, canonically conjugate to the momentum $\hat{\pi}_n$, corresponding to the displacement of $n$th ruler dipole relative to its classical equilibrium position. We choose the ruler centre-of-mass to be the origin of the reference frame. Consequently, we have the constraint
\begin{equation}
    x_{\mathrm{rCM}}=\frac{1}{N}\sum_{n=-\frac{N-1}{2}}^{\frac{N-1}{2}} x_{r,n}=0.
    \label{constrainteq1}
\end{equation}
Note that this constraint does not imply the localisation of the ruler's centre-of-mass in the lab frame. Nevertheless, upon solving the equations of motion, it is found that the velocity of the ruler centre-of-mass is identically zero in its own reference frame. As a result, the first term in Eq.~\eqref{eq:Hrfree} disappears. This constraint is effectively the same as the following constraint for the relative coordinates (for the details of this equivalence, see Appendix~\ref{app:constraints}):
\begin{equation}
    \sum_{n=-\frac{N-1}{2}}^{\frac{N-1}{2}} \phi_{n}=0.
    \label{constrainteq2}
\end{equation} 
The free Hamiltonian of the ion is
\begin{equation} \label{eq:Hionfree}
	\hat{H}_{I}= \frac{\hat{p}^2_{I}}{2M_I},
\end{equation} 
and $\hat{V}_{I r}$ describes the electromagnetic, Coulomb interaction between an assumed negatively-charged ion ($-q_I$) and the ruler through its dipoles: 
\begin{equation} \label{eq:CoulombV}
	\hat{V}_{I r} = -\frac{q_I q} {4\pi \epsilon_0} \sum_{n=-\frac{N-1}{2}}^{\frac{N-1}{2}} \left[\frac{1} {\sqrt{(w-l/2)^2+(\hat{x}_I-\hat{x}_{r,n})^2}} - \frac{1} {\sqrt{(w+l/2)^2+(\hat{x}_I-\hat{x}_{r,n})^2}} \right],
\end{equation}
where $l$ is the distance between the opposite charges $+q$ and $-q$ of a given ruler dipole (we adopt the convention $q,q_I >0$), and recall that $w$ is the fixed, perpendicular distance between the ruler dipole chain and ion; this potential results in distorting displacements of the ruler dipoles from their equilibrium positions in the presence of the ion (see later below).

We suppose that the ion is perpendicularly located sufficiently close to the ruler such that the latter must be analyzed as a discrete mass-spring lattice system (as opposed to being approximated as an elastic continuum field system), i.e., $w\lesssim a_r$. We furthermore assume that  $l, \delta\phi_n\ll w$, where $\delta\phi_n$ is the uncertainty in the $n$th dipole's displacement:
\begin{equation}
	\delta \phi_n= \sqrt{\langle\phi_{n}^2\rangle-\langle\phi_{n}\rangle^2}.
\end{equation}
 Under these conditions, the potential of Eq.\,\eqref{eq:CoulombV}  can be Taylor expanded in $\hat{\phi}_n$ to give the following simpler, approximate potential (see Appendix~\ref{app:Lagrangian} for details):  \begin{equation} \label{eq:ion-dipoleVeq}
  \hat{V}_{I r}  \approx -\frac{q_I {\mathsf{p}}_r w} {4\pi \epsilon_0} \sum_{n=-\frac{N-1}{2}} ^{\frac{N-1}{2}}\left[(\hat{x}_{I,n}^{2} + w^2)^{-3/2} + 3 \hat{\phi}_n   \hat{x}_{I,n} (\hat{x}_{I,n}^{2} + w^2)^{-5/2} \right],
\end{equation}
 where $\mathsf{p}_r = q l$ is the ruler atom electric dipole moment and $\hat{x}_{I,n} = \hat{x}_I - na_r- \hat{x}_{\mathrm{r CM}}$  is the location of the ion relative to the $n$th, rigid ruler dipole position. Constraint (\ref{constrainteq1}) gives $\hat{x}_{\mathrm{r CM}}=0$; from now on, $\hat{x}_{\mathrm{r CM}}$  will no longer appear in the equations.    The first term in Eq.~\eqref{eq:ion-dipoleVeq} describes the potential experienced by the ion due to the ruler atom dipoles in their classical equilibrium lattice positions; we henceforth define $\hat{V}_I= -\frac{q_I {\mathsf{p}}_r w} {4\pi \epsilon_0} \sum_{n=-\frac{N-1}{2}} ^\frac{N-1}{2}(\hat{x}_{I,n}^{2} + w^2)^{-3/2}$ and combine it with the free ion Hamiltonian. The total, approximate ion-ruler Hamiltonian can then be re-expressed as $\hat{H} = \hat{H}_r + \hat{H}^{\mathrm{eff}}_{I}  + \hat{V}^{\mathrm{eff}}_{I r}$, where $\hat{H}_r$ is given in Eq.~\eqref{eq:Hrfree} (without the centre of mass kinetic energy term), and the effective ion Hamiltonian and ion-ruler dipole elastic displacement interaction are, respectively, 
\begin{align}
	& \hat{H}^{\mathrm{eff}}_{I} = \frac{\hat{p}^2_{I}}{2M_I} -\frac{q_I {\mathsf{p}}_r w} {4\pi \epsilon_0} \sum_{n=-\frac{N-1}{2}} ^{\frac{N-1}{2}}(\hat{x}_{I,n}^{2} + w^2)^{-3/2} , \label{eq:IonHeff}\\
	& \hat{V}^{\mathrm{eff}}_{I r} = -\frac{3 q_I {\mathsf{p}}_r w} {4\pi \epsilon_0} \sum_{n=-\frac{N-1}{2}} ^{\frac{N-1}{2}} \hat{\phi}_n  \hat{x}_{I,n} (\hat{x}_{I,n}^{2} + w^2)^{-5/2}, \label{eq:VeffRion}
\end{align}
with $\hat{\phi}_n$  satisfying the constraint \eqref{constrainteq2}. Note that the ion-ruler Hamiltonian now depends only on the relational ion  $\hat{x}_{I,n}$ and ruler atom $\hat{\phi}_n$ dipole coordinates. 

\subsection{Transformation between local and nonlocal ruler bases}
In Appendix~\ref{app:freeruler}, we solve for the classical and quantum (Heisenberg picture) free ruler dynamics, utilizing the common approach of working in terms of the nonlocal normal mode ``position'' operator solutions:
\begin{equation}
\hat{x}_{\alpha}(t)=x_{\alpha,0}\left[\hat{a}_{\alpha}(0) e^{-i\Omega_{\alpha} t}+\hat{a}_{\alpha}^{\dag}(0)e^{i\Omega_{\alpha} t}\right],
\label{positionmodesolneq}
\end{equation}
where $x_{\alpha,0}=(\hbar/{2 m_r\Omega_{\alpha}})^{1/2}$ is the zero-point displacement uncertainty of the $\alpha$th normal mode, and the normal mode frequencies are 
\begin{equation}
\Omega_\alpha=2\omega_r\sin\left(\frac{\alpha\pi}{2N}\right), \,\alpha=1,2,\dots,N-1,
\label{modefreqeq}
\end{equation}
with $\omega_r=\sqrt{k_r/m_r}$, and $\hat{a}_{\alpha}(0)$, $\hat{a}^{\dag}_{\alpha}(0)$ are the ``phonon'' annihilation and creation operators, respectively, which satisfy the commutation relation $[\hat{a}_{\alpha}(0),\hat{a}^{\dag}_{\alpha'}(0)]=\delta_{\alpha,\alpha'}$ (with all other commutation relations vanishing). 

However, in order to understand how this system  functions as a ruler, it is necessary to also work in terms of the local, dipole atom position observables $\phi_n$ that correlate with the position observable $x_{I,n}$  of the ion system.  
To achieve this, we need to perform a change of basis from the  normal mode position operator eigenstates $|\{x_\alpha\}\rangle_r$ to the local dipole displacement operator eigenstates $|\{\phi_n\}\rangle_r$ of the ruler.  This is implemented through the following linear transformation between the mode position and dipole position coordinates:
\begin{equation}
\phi_n (t)=\sum_{\alpha=1}^{N-1} u_{\alpha,n} x_\alpha(t),
\label{normalmodes}
\end{equation}
where the $u_{\alpha, n}$ are the orthonormal mode eigenfunctions of the ruler:
\begin{equation}
    u_{\alpha, n}=\sqrt{\frac{2}{N}} \cos \left[\frac{\alpha\pi}{N}\left(n+\frac{N}{2}\right)\right],\, \alpha=1,2,\dots,N-1.
    \label{rulermodeseq}
\end{equation}
Note that we have one fewer mode ($N-1$) than the total number of ruler atom dipoles ($N$). This is because we do not include the zero frequency, $\alpha=0$ mode in the sum, which results in Eq.\,\eqref{normalmodes}  solving the constraint of Eq.\,(\ref{constrainteq2}). In other words, imposing the constraint is equivalent to simply removing the zero frequency, centre of mass mode of the ruler  (hence showing a key advantage of first working in terms of the nonlocal normal mode solutions to the free ruler equations of motion).
The inverse transformation is
\begin{equation}
x_\alpha(t)=\sum_{n=-\frac{N-3}{2}} ^{\frac{N-1}{2}}  \tilde{u}_{\alpha,n}\phi_n(t),
\label{inversetrans}
\end{equation}
where\footnote{We originally have $x_\alpha(t)=\sum_{n=-\frac{N-1}{2}} ^{\frac{N-1}{2}}  u_{\alpha,n}\phi_n(t)$, and a technical subtlety is that we need to eliminate one of the atom dipole sites in the sum in order to enforce the relational constraint equation~(\ref{constrainteq2}). Here, it does not matter which coordinate we eliminate, as long as it is not one of the ruler coordinates opposite the ion location. We choose to  eliminate the left-most ruler edge coordinate $\phi_{-\frac{N-1}{2}}=-\sum_{n=-\frac{N-3}{2}} ^{\frac{N-1}{2}} \phi_n$.} $\tilde{u}_{\alpha,n}=u_{\alpha,n}-u_{\alpha,-\frac{N-1}{2}}$.

\subsection{Second-quantized, tight binding model  of the ion-ruler system}
In this section, we will first derive an approximate, second-quantized tight binding model description of the ion-ruler system, and then obtain analytical solutions for the resulting ion-ruler quantum dynamics.
From Eqs. \eqref{positionmodesolneq} and \eqref{normalmodes}, the free ruler dipole position and momentum operators operators are respectively defined as (see Appendix~\ref{app:freeruler} for the derivation details):
\begin{equation}
\hat{\phi}_n(t)=\sum_{\alpha=1}^{N-1}x_{\alpha,0}u_{\alpha,n} \left[\hat{a}_{\alpha}(0) e^{-i\Omega_{\alpha} t}+\hat{a}_{\alpha}^{\dag}(0) e^{i\Omega_{\alpha} t}\right],
\label{rulersolneq}
\end{equation}
\begin{equation}
\hat{\pi}_n(t)=-i\sum_{\alpha=1}^{N-1} \sqrt{\frac{m_r\Omega_{\alpha}\hbar}{2}}u_{\alpha,n}\left[\hat{a}_{\alpha}(0) e^{-i\Omega_{\alpha} t}-\hat{a}_{\alpha}^{\dag}(0) e^{i\Omega_{\alpha} t}\right] ,
\label{rulermomsolneq}
\end{equation}
where the normal mode ruler frequencies are given by Eq. \eqref{modefreqeq}.  
Expressed in terms of the normal mode annihilation and creation operators, the
ruler Hamiltonian $\hat{H}_r$ takes the standard form of a sum over decoupled harmonic oscillator Hamiltonians, one for each mode:
\begin{equation}
    \hat{H}_r = \frac{1}{2}\sum_{\alpha=1} ^{N-1} \hbar\Omega_{\alpha} \left( \hat{a}_{\alpha} ^{\dag} \hat{a}_{\alpha}+ \hat{a}_{\alpha} \hat{a}_{\alpha}^{\dag}\right).
 \label{quantizedruler}
\end{equation}

Throughout this work, we restrict to situations where the ion is distant from the ruler edges labelled by $n=\pm (N-1)/2$ (with ruler dipole number $N\gg 1$). Let us first assume therefore that we initially prepare the ion in a localised state opposite the $i$th ruler dipole, where $|i|\ll (N-1)/2$. While the ion will classically remain trapped by the potential $V_I$ in the ion Hamiltonian $H_I^{\mathrm{eff}}$, quantum mechanically the ion can ``hop'' from one dipole site to the next by tunnelling through the potential barriers between the sites, leading to delocalisation of the ion\footnote{Taking into account the quantum dynamical response of the ruler dipoles (i.e., phonons) to the ion may in fact serve to localise the ion under certain conditions \cite{schmid1983,fisher1985,aslangul1985}; we do not consider such a possibility in the present work.}. However, we shall work in a parameter regime where the ion hopping timescale between nearest neighbour ruler dipoles is long compared to the timescale over which the ruler responds to the initial presence of the ion, i.e., the timescale for the ruler to measure the position of the ion. 
Restricting to $w\ll a_r$, i.e., the perpendicular distance between the ion and ruler is much smaller than the ruler dipoles' classical, static equilibrium separation $a_r$, we can then approximate $V_I$ as a delta function potential: ${w^2} (x^{2} + w^2)^{-3/2}/2\to \delta(x)$, giving  ${V}_{I}(x) \approx -q_I {\mathsf{p}}_r \delta(x-x_i)/(2\pi\epsilon_0 w)$, with $x_i=i a_r$. Neglecting for now tunnelling to the neighbouring sites, the (degenerate) ground states of the ion Hamiltonian $\hat{H}_I^{\mathrm{eff}}$ are given approximately by the bound states
\begin{equation} \label{eq:boundion}
|i\rangle_I=\sqrt{\kappa} \int dx \,e^{-\kappa |x-x_i|} |x\rangle_I,
\end{equation}
 where $\kappa={M}_I q_I \mathsf{p}_r/( 2\pi\hbar^2 \epsilon_0 w)$. Here, $|i\rangle_I$ denotes that the ion  is localised at site $i$ when the position uncertainty of the ion satisfies  $\Delta x=1/(\sqrt{2}\kappa)< a_r$.  

In order to account for the ion being localised at different sites, as well as account for tunnelling between neighbouring sites, it is convenient to adopt a second-quantized tight-binding description, where the position of the ion is expressed in terms of the number of ions at each site $n$, forming a Fock space spanned by the following Fock state basis:
\begin{equation} 
   |\psi\rangle_I=\ket{s_{-\frac{N-1}{2}}}_I \otimes \cdots \ket{s_n}_I \otimes \cdots \ket{s_{\frac{N-1}{2}}}_I.
\label{ionfockspace}
\end{equation}
If the ion is localised at site $i$, the Fock state corresponding to the single ion state $|i\rangle_I$ satisfies $s_n=\delta_{i,n}$. We then introduce creation and annihilation operators, $\hat{c}^{\dag}_n$ and $\hat{c}_n$ respectively for an ion at the $n$th site, and suppose that these operators  satisfy anticommutation relations (i.e., the ions are treated as Fermions), so that not more than one ion can occupy a given site. However, since we will be considering below only initial states that consist of a single ion, the quantum dynamics ensures that the ion number is conserved and always equal to one, and whether we treat the ions as Fermions or Bosons is then immaterial\footnote{In principle, initial states consisting of more than one ion could also be considered by working with this second quantized model. In this case, it would be necessary to also take into account the ion-ion repulsive interaction, and different quantum dynamics would result depending on whether the ions are Bosons or Fermions.}. Note that we do not in fact require the use of the full $N$-fold tensor product in Eq.\,\eqref{ionfockspace}, since we will only consider ion states that correspond to the ion being distant from the ruler edges ($|i|\ll N$), as mentioned above. In particular, including the  $\ket{s_{-\frac{N-1}{2}}}_I$  state space in Eq.\,\eqref{ionfockspace} will not give rise to any inconsistencies in the following.

The effective ion Hamiltonian~\eqref{eq:IonHeff} with rigid ruler potential becomes in the second quantized, tight-binding formulation: 
\begin{equation} 
 \hat{H}^{\mathrm{eff}}_I  = \sum_{n=-\frac{N-1}{2}} ^{\frac{N-1}{2}} \left[\nu \hat{c}_n^{\dag}\hat{c}_n +\gamma \left(\hat{c}_{n+1}^{\dag}\hat{c}_n + \hat{c}_n^{\dag}\hat{c}_{n+1} \right)\right],
\end{equation}
where the on-site binding energy $\nu$ is given by
\begin{equation}
	 \nu ={}_I\langle n| \hat{H}^{\mathrm{eff}}_I |n\rangle_I = -\frac{\hbar^2\kappa^2}{2M_I},
  \label{nueq}
\end{equation} 
and the hopping strength $\gamma$ is given by
\begin{equation}
	\gamma = {}_I\langle n+1| \hat{H}^{{\mathrm{eff}}}_I |n\rangle_I = -\frac{\hbar^2\kappa^2}{M_I} e^{-\kappa a_r} \left(\kappa a_r+1\right).
 \label{gammaeq}
\end{equation} 
With the ion wavefunction localisation condition giving $\kappa a_r>1$ (see above), we have that $e^{-\kappa a_r}\ll 1$, so that the overlap integral and hence hopping strength is only significant for nearest-neighbour sites. From Eqs.\,\eqref{nueq} and \eqref{gammaeq}, we also have that  $\gamma\ll \nu$ and from now on we neglect the hopping terms in $\hat{H}^{\mathrm{eff}}_I$. 

The ion-ruler dipole elastic displacement interaction $\hat{V}_{Ir}^{{\mathrm{eff}}}$ defined in Eq.~\eqref{eq:VeffRion} takes the following second-quantized, tight binding form:
\begin{equation} 
    \hat{V}_{Ir}^{\mathrm{eff}} = -\frac{3q_I {\mathsf{p}}_r w} {4\pi \epsilon_0} \sum_{n=-\frac{N-1}{2}} ^{\frac{N-1}{2}}  \hat{\phi}_n \hat{x}_{In} \left(\hat{x}_{In}^2 + w^2\right)^{-5/2} |_{\hat{x}_{In}=\hat{x}_I - na_r} = -\lambda \sum_{n=-\frac{N-1}{2}} ^{\frac{N-1}{2}}  \hat{c}^{\dag}_n \hat{c}_n \hat{\phi}_n,
    \label{tbvireq}
\end{equation}
where the coupling strength $\lambda$ is given by
\begin{equation} 
   \begin{split}
       \lambda&=-\frac{\langle n|\hat{V}_{Ir}^{\mathrm{eff}} |n\rangle} {\hat{\phi}_n} = \frac{3q_I {\mathsf{p}}_r w\kappa} {4\pi \epsilon_0} \int_0^{\infty} dx \,e^{-2\kappa x} x (x^2+w^2)^{-5/2}\\ 
   &= \frac{3q_I {\mathsf{p}}_r w\kappa^4} {4\pi \epsilon_0} \xi(\kappa w),
   \end{split}
\end{equation}
with $\xi(z) = \int_0^{\infty} d \tilde{x} \,
e^{-2\tilde{x}} \tilde{x} (\tilde{x}^2+z^2)^{-5/2}$,  $\tilde{x}=\kappa x$. 

The full, second quantized tight-binding form of the ion-ruler Hamiltonian is then
\begin{equation} \label{eq:tbHam}
   	\hat{H} = \sum_{n=-\frac{N-1}{2}} ^{\frac{N-1}{2}} \nu \hat{c}_n^{\dag}\hat{c}_n  + \sum_{\alpha=1} ^{N-1} \hbar\Omega_{\alpha} \hat{a}_{\alpha}^{\dag}\hat{a}_{\alpha} -\sum_{n=-\frac{N-1}{2}} ^{\frac{N-1}{2}} \sum_{\alpha=1}^{N-1}\hbar\Omega_{\alpha} \lambda_{\alpha,n}\, \hat{c}_n^{\dag}\hat{c}_n \left(\hat{a}_{\alpha} + \hat{a}_{\alpha}^{\dag}\right),
\end{equation}
where the dimensionless ion-ruler mode coupling strength is  defined as follows:
\begin{equation}
	\lambda_{\alpha,n} = \frac{\lambda x_{\alpha,0} u_{\alpha,n}}{\hbar\Omega_{\alpha}},
 \label{lambdaalphadef}
\end{equation}
with $x_{\alpha,0}=(\hbar/{2 m_r \Omega_{\alpha}})^{1/2}$, and $\Omega_{\alpha}$, $u_{\alpha,n}$ defined in Eqs. \eqref{modefreqeq} and  \eqref{rulermodeseq} respectively.

\subsection{Exact solution to the ion-ruler quantum dynamics}

We shall focus on the quantum dynamics resulting from the following example initial ion-ruler product state: 
\begin{equation} \label{eq:InitialpsiRulI}
|\Psi(0)\rangle=\frac{1}{\sqrt{2}} \left(|i_1\rangle_I+|i_2\rangle_I\right) \otimes |\psi(0)\rangle_r,
\end{equation}
where the ion is in an equal amplitude quantum superposition of stationary wavepackets centred at distinct sites $i_1$ and $i_2$, and the ruler is in the ground state of its free Hamiltonian, expressed in the non-local normal mode basis as
\begin{equation} \label{eq:rulground}
|\psi(0)\rangle_r= |0\rangle_1 \otimes|0\rangle_2 \otimes... |0\rangle_{\alpha} \otimes... |0\rangle_N.
\end{equation}
In particular, the ruler normal mode phonon occupation numbers are all initially zero.

Assuming such an ion-ruler initial product state \eqref{eq:InitialpsiRulI} is equivalent to ``plucking'' (i.e., suddenly switching on) the ion-ruler interaction at time $t=0$. Therefore, the ruler will behave as a wavelike medium with the formation of ``ripples'' that reflect from the ruler ends and do not dissipate away; the ruler would then function very poorly in terms of measuring the ion position. This problem can be addressed by inserting a switching function $S(t)$ in the interaction part of Hamiltonian \eqref{eq:tbHam}, which ensures that the interaction between the ion and the ruler is turned on sufficiently slowly:
\begin{equation} \label{eq:tbHamSwitch}
    \hat{H}=\sum_{n=-\frac{N-1}{2}} ^{\frac{N-1}{2}} \nu \hat{c}_n^{\dag}\hat{c}_n  + \sum_{\alpha=1} ^{N-1} \hbar\Omega_{\alpha} \hat{a}_{\alpha}^{\dag}\hat{a}_{\alpha}-\sum_{n=-\frac{N-1}{2}} ^{\frac{N-1}{2}} \sum_{\alpha=1}^{N-1}S(t)\hbar\Omega_{\alpha}\lambda_{\alpha,n} \hat{c}_n^{\dag}\hat{c}_n \left(\hat{a}_{\alpha} + \hat{a}_{\alpha}^{\dag}\right).
\end{equation}
Here, we model the switching function as follows:
\begin{equation} \label{eq:switchfn}
S(t)=\begin{cases} 1- e^{-t/\Delta t} & t> 0\\ 0 & t<0\end{cases}.
\end{equation}
This switching function equals zero for $t <0$ and approaches one as $t\rightarrow \infty$, switching on at around $t= 0$ over the duration $\Delta t$; in the limit $\Delta t\rightarrow 0$, the switching function \eqref{eq:switchfn} coincides with the step function $\Theta(t)$. 

Hamiltonian \eqref{eq:tbHamSwitch}, which neglects the tunnelling of the ion between neighbouring sites, resembles that for an optomechanical many-body system comprising multiple cavity modes and multiple mechanical modes \cite{bruschi2019time}. The fact that the ion-ruler interaction Hamiltonian commutes with the on-site ion Hamiltonian allows for the quantum dynamics to be solved analytically in closed form; we apply the method of analysis given in Ref.~\cite{bruschi2019time} to express the unitary time-evolution operator $\hat{U}(t)$,  denoted in Sec. \ref{sec:ToyMeas} as $\hat{U}_{Sr}$, in the following manner:
\begin{equation}
     \hat{U}(t)=e^{-i\frac{\nu}{\hbar} \sum_n \hat{c}_n^{\dag}\hat{c}_n t}\, e^{-i \sum_{\alpha,n} f(\alpha, n, t) (\hat{c}_n^{\dag}\hat{c}_n)^2 } \,e^{\sum_{\alpha,n} \left[g(\alpha, n, t) \hat{a}_{\alpha}^{\dag} -g^*(\alpha, n, t) \hat{a}_{\alpha} \right] c_n^{\dag}c_n}\, e^{-i\sum_\alpha\Omega_\alpha t \hat{a}_{\alpha}^{\dag}\hat{a}_{\alpha}},
 \label{unitaryeq}
\end{equation}
where $f = (F_1 +F_2 F_3)$, $g = (F_3 -iF_2) e^{-i\Omega_{\alpha}t}$, and $F_1$, $F_2$, and $F_3$ are time-dependent functions defined respectively as 
\begin{equation}
	\begin{split} \label{eq:timedepen}
		& F_1(\alpha,n, t)=-2\lambda^2_{\alpha,n}\int_0^{\Omega_{\alpha} t} d\tau \left(1-e^{-\frac{\tau}{\Omega_{\alpha} \Delta t}}\right) \sin\left( \tau\right)\int_0^{\tau} d\tau^{\prime} \left(1-e^{-\frac{\tau^{\prime}}{\Omega_{\alpha} \Delta t}}\right) \cos\left(\tau^{\prime} \right),\\
		& F_2(\alpha,n, t)=-\lambda_{\alpha,n}\int_0^{\Omega_{\alpha} t} d\tau \left(1-e^{-\frac{\tau}{\Omega_{\alpha} \Delta t}}\right) \cos\left(\tau\right),\\
		& F_3(\alpha,n, t)=-\lambda_{\alpha,n}\int_0^{\Omega_{\alpha} t} d\tau \left(1-e^{-\frac{\tau}{\Omega_{\alpha} \Delta t}}\right) \sin\left(\tau\right).
	\end{split}
\end{equation}
 Cross term contributions of the form $e^{-i\sum_{\alpha, n\neq m}f(\alpha,n,m,t) (\hat{c}_n^{\dag}\hat{c}_n \hat{c}_m^{\dag}\hat{c}_m)}$ are neglected in the unitary operator expression \eqref{unitaryeq}, since they  are only relevant for states describing more than one ion occupying different sites; here we restrict ourselves to single ion states.

The initial state $|\Psi(0)\rangle$ given by Eq.\,\eqref{eq:InitialpsiRulI} evolves into an entangled state between the ion and the ruler:
\begin{equation} \label{eq:IonRulerTimet}
|\Psi(t)\rangle =\hat{U}(t)|\Psi(0)\rangle=\frac{1}{\sqrt{2}} e^{-i\frac{\nu}{\hbar} t} \left[ \ket{i_1}_I \ket{\Phi_1(t)}_r + \ket{i_2}_I \ket{\Phi_2(t)}_r \right],
\end{equation}
where $\ket{\Phi_m(t)}_r= \prod_{\alpha=1}^{N-1} e^{-i f(\alpha,i_m, t)} | g(\alpha, i_m, t) \rangle_r$, $m=1, 2$,  with $\ket{g(\alpha, i_m, t)}_r$ a coherent state of normal mode $\alpha$. From Eq.\,\eqref{eq:IonRulerTimet}, the density matrix of the ion-ruler system is 
\begin{equation} \label{eq:IonRulerdensityTimet}
\hat{\rho}(t)=|\Psi(t)\rangle \langle\Psi(t)|= \frac{1}{2} \sum_{m,m^\prime=1}^2 \ket{i_m}_I \bra{i_{m^\prime}} \otimes \ket{\Phi_m}_r \bra{\Phi_{m^\prime}}.
\end{equation}

A necessary condition for the ruler to measure the position of the ion is that the off-diagonal terms of the reduced density matrix of the ion subsystem in its site position basis become suppressed over time, i.e., the ruler decoheres the initial ion superposition state. 
We obtain for the off-diagonal term of the ion reduced density matrix
\begin{equation} \label{eq:rhoOffDiag}
	\begin{split}
		\rho_I^{i_1 i_2}(t)&= {}_I \bra{i_1} \mathrm{Tr}_r\left[ \hat{\rho}(t)  \right] \ket{i_2}_I=\\
		&=\frac{1}{2} \mathrm{Tr}_r\left\{ \prod_{\alpha, \beta =1}^{N-1} e^{-i \left[f(\alpha,i_1, t)-f(\beta,i_2, t)\right]} \ket{g(\alpha, i_1, t)}_r \bra{g(\beta, i_2, t)} \right\}.
	\end{split}
\end{equation}
Recalling the formula for the inner product between two coherent states $\ket{a}$ and $\ket{b}$, $\langle b| a \rangle=\exp\left[-1/2(|b|^2 + |a|^2 -2b^\ast a )\right]$, from Eqs.~\eqref{eq:timedepen} and \eqref{eq:rhoOffDiag}  we obtain for the ion coherence 
\begin{equation} \label{syscohere1}
    \begin{split}
       {C}_I(t)&=2|{\rho}_I^{i_1 i_2}(t)|= \prod_{\alpha, \beta =1}^{N-1} |\braket{g(\alpha, i_1, t)| g(\beta, i_{2}, t)}|=\\
        &= \prod_{\alpha, \beta =1}^{N-1} \delta_{\alpha,\beta} \exp\left\{-\frac{1}{2} \mathrm{Re} \left[ 
|g(\alpha, i_1, t)|^2 + |g(\beta, i_2, t)|^2 -2g^\ast(\alpha, i_1, t) g(\beta, i_2, t)\right] \right\}=\\
        &=\exp\left\{-\frac{1}{2}\sum_{\alpha=1}^{N-1} \left[ (F_3(\alpha,i_1, t)-F_3(\alpha,i_2, t))^2 + (F_2(\alpha,i_1, t)-F_2(\alpha,i_2, t))^2\right] \right\}.
    \end{split}
\end{equation}
For times longer than the switch-on duration $\Delta t$, the functions $F_2$ and $F_3$ become approximately 
\begin{align}
F_2(\alpha,n, t)&\approx-{\lambda_{\alpha,n}}\left(\sin(\Omega_\alpha t)-\frac{\Omega_{\alpha}\Delta t}{1+\Omega_\alpha^2\Delta t^2} \right),\label{f2eq}\\
F_3 (\alpha,n, t)&\approx{\lambda_{\alpha,n}}\left(\cos(\Omega_\alpha t)-\frac{1}{1+\Omega_\alpha^2\Delta t^2} \right),
\label{f3eq}
\end{align}
where $\lambda_{\alpha,n}$ is defined in Eq. \eqref{lambdaalphadef}.
Provided the switch-on duration satisfies  $\Delta t \gg 1/\Omega_{\alpha=1}$,
expressions \eqref{f2eq} and \eqref{f3eq} can then be further simplified to 
\begin{equation}\label{eq:Flongtime}
	F_2 (\alpha, n, t)\approx-{\lambda_{\alpha,n}} \sin(\Omega_\alpha t),\qquad
	F_3 (\alpha, n, t)\approx{\lambda_{\alpha,n}} \cos(\Omega_\alpha t).
\end{equation}
For $N\gg 1$, the above condition on the switch-on duration can be rewritten as $\Delta t\gg L/(\pi c_r)$, where $L=(N-1)a_r\approx Na_r$ is the classical, equilibrium free ruler length and $c_r=\omega_r a_r$ is the ruler elastic wave propagation speed in the long wavelength (equivalently low frequency) limit. In particular, the ion-ruler coupling is switched on more slowly than the time for an acoustic wave to propagate the length of the ruler. 
For $t>\Delta t$, the ion coherence \eqref{syscohere1} is then given approximately by the following long time limit expression 
\begin{equation} \label{eq:coherenceTinfty}
	\lim_{t\rightarrow\infty}{C}_I(t) = \exp\left\{-\sum_{\alpha=1}^{N-1} \frac{\lambda^2}{2\hbar N m_r\Omega_{\alpha}^3} \left[\cos\left(\frac{\alpha\pi}{N}{\left(i_1+\frac{N}{2}\right)}\right)-\cos\left(\frac{\alpha\pi}{N}{\left(i_2+\frac{N}{2}\right)}\right) \right]^2 \right\}.
\end{equation}
Note that Eq. \eqref{eq:coherenceTinfty} is time-independent.

%\begin{figure}
%\centering
%\begin{subfigure}{.5\textwidth}
%  \centering \includegraphics[width=1\linewidth]{coherence1.pdf}
%  \caption{}
%  \label{coherence1}
%\end{subfigure}
%\begin{subfigure}{.5\textwidth}
%\centering
%\includegraphics[width=1\linewidth]{coherence2.pdf} 
%\caption{}
%\label{coherence2}
%\end{subfigure}
%\caption{\label{coherence}Ion coherence \eqref{eq:coherenceTinfty} versus ruler dipole number $N$ and separation $|i_1-i_2|$ between ion superposition states; the ruler dipole parameter units are $m_r=k_r=\hbar=1$. (a) Coupling strength $\lambda=0.3$. (b) Coupling strength $\lambda=2$ and scaling $m_r \to N m_r$, $k_r \to N k_r$.}
%\end{figure}
\begin{figure}[htbp]
    \centering
    \begin{subfigure}{0.49\linewidth}
        \centering
        \includegraphics[width=\linewidth]{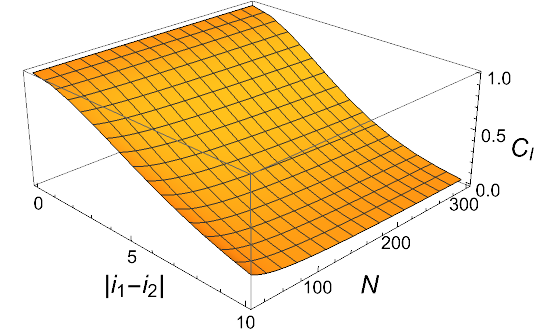}
        \caption{}
        \label{coherence1}
    \end{subfigure}\hfill
    \begin{subfigure}{0.49\linewidth}
        \centering
        \includegraphics[width=\linewidth]{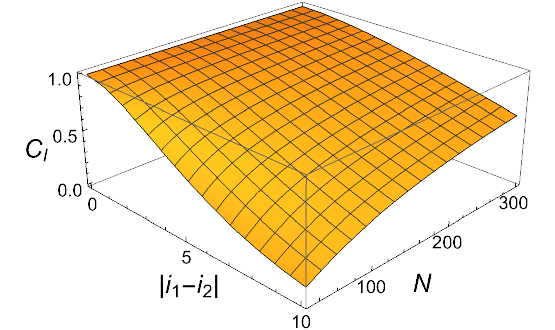}
        \caption{}
        \label{coherence2}
    \end{subfigure}
    \caption{Ion coherence \eqref{eq:coherenceTinfty} versus ruler dipole number $N$ and separation $|i_1-i_2|$ between ion superposition states; the ruler dipole parameter units are $m_r=k_r=\hbar=1$. (a) Coupling strength $\lambda=0.3$. (b) Coupling strength $\lambda=2$ and scaling $m_r \to N m_r$, $k_r \to N k_r$.}
    \label{coherence}
\end{figure}

Fig.~\ref{coherence} plots the dependence of the long-time limit coherence  given by Eq.~\eqref{eq:coherenceTinfty} on both the ion superposition separation $|i_1 - i_2|$ and the ruler dipole number $N$, for some example ion-ruler system parameters. As might be expected from enviromentally induced decoherence (with the  ruler acting as an environment for the ion), the coherence becomes smaller the larger the ion superposition separation $|i_1 - i_2|$, as can be seen from Fig.~\ref{coherence1} by fixing a given value for $N$ and looking at the dependence on the separation. 
Less expected in Fig.~\ref{coherence1} is a weaker, but still progressive decrease in coherence with increasing ruler dipole number $N$ (equivalently increasing ruler length) for a fixed given ion superposition separation $|i_1 - i_2|$; we might have expected that the fixed separation coherence would not depend on ruler length when the latter is much larger than the former (recall we are assuming that the ion is distant from the ruler edges). The resolution lies in the fact that we are considering a one-dimensional mass-spring model of a ruler, which in contrast to a more realistic two or three-dimensional model,  gives rise to infra-red type signatures (i.e., boundary size effects) in local properties. In particular, as our one-dimensional ruler becomes longer, it gets more ``floppy'', and the zero-point fluctuations in the dipole displacements $\delta\phi_n$ grow. Such fluctuations will cause dephasing and hence a progressive decrease in the ion coherence \eqref{eq:coherenceTinfty}  with increasing $N$ as seen in Fig.~\ref{coherence1}. Note that such dephasing due to ruler zero-point fluctuations is not the same as decoherence due to the ruler becoming entangled with the ion (and measuring the position of the latter); both decoherence and dephasing result in a reduction of the ion coherence \eqref{eq:coherenceTinfty} that cannot be distinguished without measuring the ruler response to the ion as well (as discussed in Sec. \ref{sec:measurement} below).      

The above-described ruler length signature can be avoided by scaling both the ruler atom mass $m_r$ and spring constant $k_r$ by the factor $N^s$, with $s>0$ some scaling exponent, i.e., by making the ruler progressively stiffer and correspondingly more massive as its length increases, hence capturing some distinguishing scaling aspects of longitudinal vibrational modes in two and three-dimensional rulers. In Fig.~\ref{coherence2}, we show the ion coherence for the scalings $m_r \to N m_r$ and $k_r \to N k_r$ (i.e., $s=1$), which effectively corresponds to a two-dimensional model. Since including the scaling makes the ruler stiffer, we have correspondingly increased the ion-ruler coupling strength from $\lambda=0.3$ to $\lambda=2$ in order to increase the local distortion of the ruler opposite to the ion location. 
Note that the coherence now in fact increases with ruler length and fixed ion superposition separation, in contrast to when there is no scaling; this is because the dipole zero-point fluctuations decrease with increasing dipole mass and spring stiffness, resulting in less dephasing.

\section{The quantum ruler as a position measurement device}
\label{sec:measurement}
In the previous section, we examined the reduced ion system state in the long time limit, tracing out the state of the ruler. This gave us some indirect information about the behaviour of the ruler  as a quantum, many degree of freedom environment interacting with the ion. In this section, we investigate the behaviour of the ruler as a quantum measuring device for the ion that is initially in a superposition state \eqref{eq:InitialpsiRulI} and with the ruler initially in its ground state \eqref{eq:rulground}. We shall first obtain the ion-ruler density matrix in the local position representation $|\{\phi_n\}\rangle_r$ of the ruler dipoles, and investigate the ruler response to the ion  by tracing out the latter and considering the average $\langle\phi_n\rangle$ and variance $\delta\phi_n$ of the ruler dipole displacements versus dipole site $n$. We then selectively trace out the ruler dipoles, except for those directly opposite the ion, giving a reduced density matrix for the ion and two nearest dipoles whose displacements $\phi_{i_{1(2)}}$ respond to the ion's local presence. The resulting reduced density matrix is then compared to that which assumes an initial mixed state for the ions through a certain joint measurement of the ion and ruler dipole displacements, in order to quantify the extent to which the extended material ruler acts as a quantum position measuring device.      

\subsection{Local description of the ruler dipoles state}
We first express the density operator \eqref{eq:IonRulerdensityTimet} in the nonlocal mode position representation $|\{x_\alpha\}\rangle_r$ as follows:
\begin{equation}\label{xdensityeq}
	\begin{split}
		\hat{\rho}(t)=&\prod_{\alpha,\beta=1}^{N-1}\int dx_\alpha d\tilde{x}_\beta |x_\alpha\rangle_r \langle x_\alpha| \Psi(t)\rangle \langle\Psi(t)|\tilde{x}_\beta\rangle_r \langle \tilde{x}_\beta|\\
		=&\frac{1}{2}\sum_{m,m^\prime=1}^2\prod_{\alpha,\beta=1}^{N-1}\int dx_\alpha d\tilde{x}_\beta \psi_{i_m} (x_\alpha,t)\psi_{i_{m^\prime}}^\ast (\tilde{x}_\beta,t) |i_m\rangle_I \langle i_{m^\prime}| \otimes |x_\alpha\rangle_r \langle  \tilde{x}_\beta|,
	\end{split}
\end{equation}
where the wave function $\psi_{i_m}(x_\alpha,t)$ of mode $\alpha$ is defined as $\psi_{i_m}(x_\alpha,t)={}_r \langle x_\alpha|\Phi_m(t)\rangle_r$. 
From Eq.\,\eqref{eq:Flongtime} and the definition for $\ket{\Phi_m(t)}_r$ given just below \eqref{eq:IonRulerTimet}, we have in the long time limit, $\ket{\Phi_m}_r= \prod_{\alpha=1}^{N-1} e^{-i f(\alpha,i_m,t)} |{{\lambda} _{\alpha,i_m}} \rangle_r$, and the mode $\alpha$ wave function is
\begin{equation}
\psi_{n}(x_\alpha,t)= \frac{1}{\sqrt{\sqrt{2\pi}x_{\alpha,0}}} e^{-i f(\alpha,n,t)}e^{-\frac{1}{4}\left(\frac{x_\alpha} {x_{\alpha,0}}-2{\lambda} _{\alpha,n}\right)^2},
\label{coherordinate}
\end{equation}
where $f=F_1+F_2F_3$ [with the $F_i$ functions defined in Eq.\,\eqref{eq:Flongtime}], $x_{\alpha,0}=(\hbar/{2 m_r\Omega_{\alpha}})^{1/2}$, and  ${\lambda} _{\alpha,n}=\lambda x_{\alpha,0} u_{\alpha,n}/(\hbar\Omega_{\alpha})$ [see Eq.\,\eqref{lambdaalphadef}].

We next express the ion-ruler system density operator \eqref{xdensityeq} in terms of the local, ruler dipole displacement coordinate representation $\ket{\{\phi_n\}}_r$, by inserting on the left and right sides the following resolution of the identity\footnote{Note that $n$ ranges from $-(N-3)/2$ to $(N-1)/2$, since we have eliminated the $\phi_{(N-1)/2}$ coordinate by expressing it in terms of the remaining $\phi_n$s through the constraint \eqref{constrainteq2}.}:
\begin{equation}
\mathbb{1}=\sqrt{N}\prod_{n=-\frac{N-3}{2}} ^{\frac{N-1}{2}}\int d\phi_n|\phi_{n}\rangle_r \langle \phi_{n}|.
\end{equation}
The overall $\sqrt{N}$ normalization factor is the determinant of the $N-1$ dimensional Jacobian matrix: $\frac{\partial\left(\{x_{\beta}\}\right)}{\partial\left(\{\phi_n\}\right)}=\sqrt{N}$. Integrating over the nonlocal mode coordinates $x_{\alpha}$, $\tilde{x}_{\alpha}$, and
 using the fact that ${}_r\langle\{\phi_n\}|x_\alpha \rangle_r =\delta(x_\alpha-\sum_n \tilde{u}_{\alpha,n} \phi_n)$ [see Eq. \eqref{inversetrans}],  the density operator \eqref{xdensityeq} becomes
\begin{equation}\label{phidensityeq}
\hat{\rho}(t)= \frac{\sqrt{N}}{2}\sum_{m,m'=1}^2\prod_{i,j=-\frac{N-3}{2}} ^{\frac{N-1}{2}}\int d\phi_id\phi_j^{\prime} \prod_{\alpha=1}^{N-1} \tilde{\psi}_{i_m} (\{\phi_n\}, t)\tilde{\psi}^\ast_{i_{m^\prime}} (\{\phi_n\}, t) |i_m\rangle_I \langle i_{m^\prime}|\otimes |\phi_{i}\rangle_r \langle \phi_{j}^{\prime}|,
\end{equation}
where $\tilde{\psi}_{i_m} (\{\phi_n\}, t) = \psi_{i_m} \left(\sum_n \tilde{u}_{\alpha,n} \phi_n,\,t\right)$.

We now give the expression for the reduced density operator of the ion and the dipoles located at the ruler sites $i_1$, $i_2$. This amounts to tracing the density operator \eqref{phidensityeq} over the dipoles at all of the other sites $n\neq {i_1}, \,{i_2}$. We obtain
\begin{equation}
	\begin{split}
		\hat{\rho}_\mathrm{Ir}(t)&=\mathrm{Tr}_{{\phi}_{n\neq i_1,i_2}}\left[\hat{\rho}(t)\right]=\\
&=\int d\phi_{i_1} d\phi_{i_1}^{\prime}d\phi_{i_2} d\phi_{i_2}^{\prime}|\phi_{i_1}\rangle_r \langle \phi_{i_1}^{\prime}| \otimes |\phi_{i_2}\rangle_r \langle \phi_{i_2}^{\prime} | \prod_{n\neq i_1,i_2}\int d\phi_n\,  {{}_r\langle}\{\phi_n\}|\hat{\rho}(t)|\{\phi_n\}\rangle_r,
\label{rhored}
	\end{split}
\end{equation}
where now the reduced state lives on the tensor product of the Hilbert spaces of three subsystems: the ion and the ruler dipoles at sites $i_1$ and $i_2$. The elements of this density operator are given explicitly in Appendix~\ref{app:localrho}.

\subsection{Ruler response to the ion}
In this section, we determine how the ruler responds to the ion in the long time limit. Consider first the situation where the ion is localised at a single site $i$ [i.e., in a bound state \eqref{eq:boundion}]  and the ruler initially in its ground state \eqref{eq:rulground}; using the long time solution $\ket{\Phi(t)}_r= \prod_{\alpha=1}^{N-1} e^{-i f(\alpha,i,t)} | {{\lambda} _{\alpha,i}} \rangle_r$, with $\phi_n=\sum_{\alpha=1}^{N-1} x_\alpha u_{\alpha,n}$, we obtain respectively for the average ruler dipole $n$ coordinate displacement and the uncertainty in the latter: 
\begin{equation}
\langle\phi_n\rangle=\lim_{t\rightarrow\infty}{\mathrm{Tr}}\left[\hat{\phi}_n\hat{\rho}(t)\right] =2\sum_{\alpha=1}^{N-1} u_{\alpha,n} x_{\alpha,0} {\lambda}_{\alpha,i}
\label{avephieq}
\end{equation}
and 
\begin{equation}
    \delta \phi_n= \sqrt{\langle\phi_{n}^2\rangle-\langle\phi_{n}\rangle^2}=\sqrt{\sum_{\alpha=1}^{N-1} u_{\alpha,n}^2x_{\alpha,0}^2},
    \label{deltaphieq}
\end{equation}
where $u_{\alpha,n}$ is defined in Eq.\,\eqref{rulermodeseq}, $x_{\alpha,0}=(\hbar/{2 m_r\Omega_{\alpha}})^{1/2}$, and ${\lambda}_{\alpha,i}=\lambda x_{\alpha,0} u_{\alpha,i}/(\hbar\Omega_{\alpha})$, with $\Omega_{\alpha}$ defined in Eq.\,\eqref{modefreqeq}. 
Note that the ruler dipole displacement uncertainty \eqref{deltaphieq} does not depend on the interaction with the ion (i.e., no dependence on the coupling strength $\lambda$), coinciding with the free ruler quantum zero-point uncertainty in its ground state.

\begin{figure}[htbp]
    \centering
    \begin{subfigure}{0.49\linewidth}
        \centering
        \includegraphics[width=\linewidth]{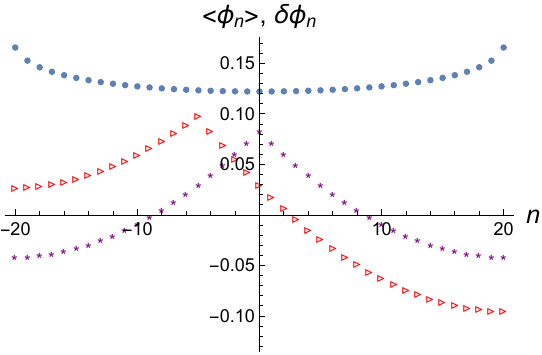}
        \caption{}
        \label{average1a}
    \end{subfigure}\hfill
    \begin{subfigure}{0.49\linewidth}
        \centering
        \includegraphics[width=\linewidth]{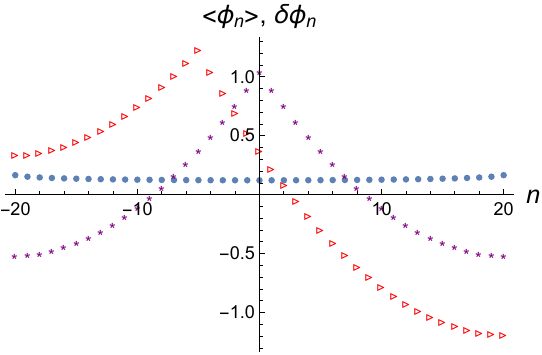}
        \caption{}
        \label{average1b}
    \end{subfigure}
    \caption{Long time limit average ruler dipole displacement $\langle\phi_n\rangle$ versus dipole site label $n$ for an ion localized at $i=0$ (purple stars) and $i=-5$ (red triangles), and ion-ruler coupling strengths (a) $\lambda=2$; (b) $\lambda=25$. The dipole displacement uncertainty $\delta\phi_n$ is also shown for comparison (blue dots). The other parameters used are $N=41$, $k_r=m_r=41$, and $\hbar=1$.}
    \label{precision1}
\end{figure}

In Fig.\,\ref{precision1}, we plot the average ruler dipole displacement $\langle\phi_n\rangle$ and uncertainty $\delta\phi_n$ versus dipole label $n$ for an example ruler length $N=41$, and two different ion-ruler coupling strengths: $\lambda=2$ (Fig. \ref{average1a}) and $\lambda=25$ (Fig.\,\ref{average1b}). The ruler dipole mass and spring constant are scaled respectively as $m_r=Nm_{r0}=41$ and $k_r=Nk_{r0}=41$, with dipole parameter units $m_{r0}=k_{r0}=\hbar=1$.

From Fig. \ref{precision1}, we see that the ruler dipole displacement   $\langle\phi_n\rangle$ is a local maximum where the ion is localised at the example sites $i=0$ and $i=-5$. This is as we require in order to have the ruler locate the ion. However, the ruler response is non-local, with the dipoles having non-negligible displacement magnitudes all the way to the edges of the ruler at $n=\pm (N-1)/2$. As mentioned in the beginning of Sec. \ref{sec:quantumruler}, this non-local ruler response is a consequence of the one-dimensional mass-spring nature of the ruler model; if we were to instead use a more realistic two or three-dimensional, extended mass-spring model of a material ruler, then the dipoles would be most displaced in the neighbourhood of the ion's location, with the displacement amplitudes decaying away in magnitude as we move away from the ion location, hence given a more desirable localised response\footnote{A price to pay, however, would be a  more complicated normal vibrational mode analysis of the extended two or three dimensional mass-spring structures.}. 

Another way to understand the nonlocal ruler response seen in Fig.\,\ref{precision1} is as a consequence of the relative coordinate constraint \eqref{constrainteq2}: $\sum_{n=-(N-1)/2}^{(N-1)/2}\langle\phi_n\rangle=0$. In particular, since there is a segment of the ruler where $\langle\phi_n\rangle>1$, we must also have complementary segments where $\langle\phi_n\rangle<1$, such that the negative region ``areas'' of the $\langle\phi_n\rangle$ versus $n$ curve cancel the positive region ``area''; we have verified that $\sum_{n=-(N-1)/2}^{(N-1)/2}\langle\phi_n\rangle=0$, as must follow from Eq. \eqref{constrainteq2}.

 In Fig.\,\ref{precision1}, we note that the dipole displacement uncertainties $\delta\phi_n$ increase towards the ruler edges. This is again a feature of the one dimensional nature of the mass-spring ruler model, where the dipoles become progressively more ``floppy'' with decreasing effective spring constants, the closer they are located to the ruler edges. With our example choice of coupling strength $\lambda=2$ (Fig.\,\ref{average1a}), we have  $\langle\phi_i\rangle < \delta\phi_i$, i.e., the average local dipole displacement at the ion location $i$ is smaller than the dipole zero-point uncertainty there; this corresponds to the ``weak'' measurement regime, where a large (ensemble) number of repeated measurements of the ruler dipole displacements is required in order to accurately determine the ion location. On the other hand, for the example choice of coupling strength $\lambda=25$ (Fig.\,\ref{average1b}), we have  $\langle\phi_i\rangle \gg \delta\phi_i$, i.e., the average local dipole displacement at the ion location $i$ is much larger than the dipole zero-point uncertainty there; this corresponds to the ``strong'' measurement regime, where we can accurately determine the ion location without a large (ensemble) number of repeated measurements on the ruler dipole displacements. 

Returning to the situation where the ion is in a superposition of two states localised at distinct sites $i_1$ and $i_2$, with the ruler initially in the ground state of its free Hamiltonian [Eq.\,\eqref{eq:InitialpsiRulI}], we obtain respectively for the ruler dipole $n$ average displacement and displacement uncertainty in the long time limit:   
\begin{align}
\langle\phi_n\rangle& =\sum_{\alpha=1}^{N-1} u_{\alpha,n} x_{\alpha,0} \left(\lambda_{\alpha,i_1}+\lambda_{\alpha,i_2}\right),\label{avephi2eq}\\
\delta \phi_n &=\sqrt{\sum_{\alpha=1}^{N-1} u_{\alpha,n}^2x_{\alpha,0}^2+ \left[\sum_{\alpha=1}^{N-1}u_{\alpha,n}x_{\alpha,0}\left(\lambda_{\alpha,i_1}-\lambda_{\alpha,i_2}\right)\right]^2}.\label{uncert2eq}
\end{align}

Note that, in contrast to the ruler dipole displacement uncertainty for the ion localised at a single site considered above, the displacement uncertainty now depends on the interaction between the ion and the ruler (characterised by the coupling strength $\lambda$). 

\begin{figure}[htbp]
    \centering
    \begin{subfigure}{0.49\linewidth}
        \centering
        \includegraphics[width=\linewidth]{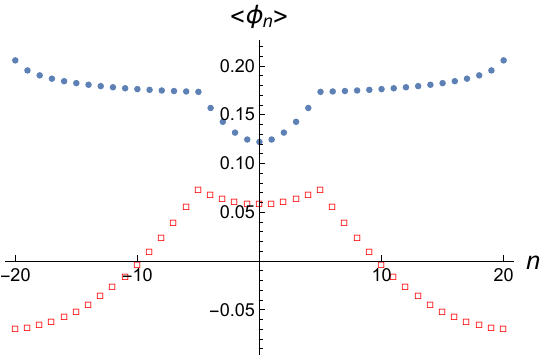}
        \caption{}
        \label{average2a}
    \end{subfigure}\hfill
    \begin{subfigure}{0.49\linewidth}
        \centering
        \includegraphics[width=\linewidth]{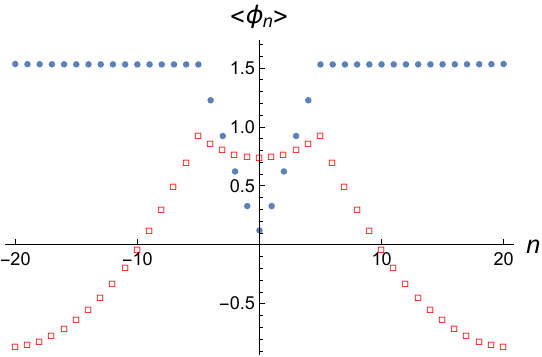}
        \caption{}
        \label{average2b}
    \end{subfigure}
    \caption{Long time limit average ruler dipole displacement $\langle\phi_n\rangle$ (red squares) and dipole displacement uncertainty $\delta\phi_n$ (blue dots) versus dipole site label $n$ for an ion in a superposition state with $i_1=-5$ and $i_2=5$. The ion-ruler coupling strengths are (a) $\lambda=2$; (b) $\lambda=25$. The other parameters used are $N=41$, $\lambda=25$, $k_r=m_r=41$, and $\hbar=1$.}
    \label{precision2}
\end{figure}

 Fig.\,\ref{precision2} plots the average ruler dipole displacement  $\langle\phi_n\rangle$ and uncertainty $\delta\phi_n$ versus dipole site label $n$ for the same example parameters as in the single site localisation situation considered above, and with superposition sites $i_1=-5$ and $i_2=5$. From Fig. \ref{precision2}, we see that the ruler dipole displacement $\langle\phi_n\rangle$ is a local maximum where the ion is localised at the sites $i_1=-5$ and $i_2=5$ in the superposition. One may also verify that the average dipole displacements satisfy the relative coordinate constraint \eqref{constrainteq2}: $\sum_{n=-(N-1)/2}^{(N-1)/2}\langle\phi_n\rangle=0$.  However, the ion-ruler $\lambda$ coupling-dependent contribution to the uncertainty [second term in the square root expression \eqref{uncert2eq}] dominates over the free ruler zero-point uncertainty [first term in the square root expression \eqref{uncert2eq}], and in fact is larger than the local maximum average dipole displacements $\langle\phi_{i_{1(2)}}\rangle$, independently of the selected coupling strength $\lambda$. This implies that, even for a large ion-ruler coupling strength, a much larger (ensemble) number of repeated measurements on the ruler dipole displacements are required in order to accurately determine the ion locations in a superposition (or mixture) of localised site states than for the situation where the ion is in a single site localised state. From Fig. \ref{average2b}, we see that  the large, $\lambda$ coupling-dependent uncertainty magnitude $\delta\phi_n$ extends to the ends of the ruler at an approximately constant value, and dips sharply between $i_1(=-5)<n<i_2(=5)$. This non-local, $\lambda$-dependent uncertainty is again a consequence of the one-dimensional nature of the ruler model; for a two or three-dimensional mass-spring model of an extended material ruler, we would expect the uncertainty to be more localised in the neighbourhood of the localised ion positions in the considered superposition state. 

\subsection{Quantum measurement scheme for superpositions of positions}

While the ruler exhibits a large quantum uncertainty in its dipole displacements for a strongly coupled  ion that is initially in a superposition of localised site states $i_1$ and $i_2$ (see Fig. \ref{precision2}), there is no difference between these $\langle \varphi_n\rangle$ and $\delta\varphi_n$ outcomes and those for an equal mixture of corresponding ion position states. Consequently, it is not possible to distinguish such a state by measurements of the ruler alone from alternatively having the ion in a mixture of the localised site states $i_1$ and $i_2$; as discussed in Sec. \ref{sec:ToyMeas}, we necessarily require a joint measurement that acts on both the ion system and ruler. 
We define such a joint quantum measurement through the projector $\hat{\Pi}^* = \ket{\Psi^*}\bra{\Psi^*}$, where 
\begin{eqnarray}
|\Psi^{\ast}\rangle=\frac{1}{\sqrt{2}} \left[ |i_1\rangle_I \ket{\bar{\chi}_{1|1}}_r \ket{\bar{\chi}_{2|1}}_r + |i_2\rangle_I \ket{\bar{\chi}_{1|2}}_r\ket{\bar{\chi}_{2|2}}_r \right],
\label{measurement}
\end{eqnarray}
The state $\ket{\bar{\chi}_{l|m}}$ for $l,m =1,2$ corresponds to the distortion of the ruler dipoles $l$ when the ion is initially localised at site $i_m$. The measurement projector $\hat{\Pi}^*$ is analogous to the measurement $M_+$ defined in Eq.\,\eqref{eq:Mpm} in Sec.\,\ref{sec:ToyMeas}. The explicit form of the states $\ket{\bar{\chi}_{l|m}}$ is
\begin{equation}
    \ket{\bar{\chi}_{l|m}}=\frac{1}{\sqrt{2c\delta\phi_{i_l}}}\int_{\langle\phi_{i_l}\rangle_m-c\delta\phi_{i_l}}^{\langle\phi_{i_l}\rangle_m+c\delta\phi_{i_l}} d\phi_{i_l} \ket{\phi_{i_l}}, ~ l,m=1,2.
    \label{chiket}
\end{equation}
Here, $\langle\phi_{i_l}\rangle_{m}$ is the average displacement of the $i_l$-th ruler site when the ion is localised at $i_m$; for the case $i_1=-i_2$, we have from Eq. \eqref{avephi2eq}:
\begin{eqnarray}
\langle\phi_{i_1}\rangle_1&=&\langle\phi_{i_2}\rangle_2=2\sum_{\alpha=1}^{N-1}u_{\alpha,i_1} x_{\alpha,0} \lambda_{\alpha,i_1},\cr 
\langle\phi_{i_2}\rangle_1&=&\langle\phi_{i_1}\rangle_2=2\sum_{\alpha=1}^{N-1}u_{\alpha,i_1} x_{\alpha,0} \lambda_{\alpha,i_2}. 
\end{eqnarray} 
The integration range in the definition \eqref{chiket} for the state $\ket{\bar{\chi}_{l|m}}$ reflects the precision of the dipole displacement measurement, with $c$ an adjustable ``precision'' parameter and with the scale set by $\delta\phi_{i_l}$, the uncertainty in the free ruler dipole displacement at the ion site $i_l$ [see Eq. \eqref{deltaphieq}]:
\begin{equation}
    \delta \phi_{i_l}=\sqrt{\sum_{\alpha=1}^{N-1} u_{\alpha,i_l}^2x_{\alpha,0}^2}.
    \label{deltaphi2eq}
\end{equation}

We then define the ion-ruler joint measurement coherence~\cite{baumgratz2014quantifying} as follows:
\begin{equation}
C^*=\frac{{\mathrm{Tr}} \left[  \hat{\Pi}^*\hat{\rho}_\mathrm{pure} \right]-{\mathrm{Tr}} \left[ \hat{\Pi}^*\hat{\rho}_\mathrm{mix} \right]}{{\mathrm{Tr}} \left[  \hat{\Pi}^*\hat{\rho}_\mathrm{mix} \right]},
\label{ratio1}
\end{equation}
where $\hat{\rho}_\mathrm{pure}$ denotes  the ion-ruler density operator in the long time limit, with the ion initially prepared in a pure superposition state and the ruler initially in its free ground state, while $\hat{\rho}_\mathrm{mix}$ denotes the ion-ruler density operator in the long-time limit, with the ion initially prepared in a mixed state and the ruler initially in its free ground state. 
The coherence $C^*$  quantifies the extent to which the joint measurement that we have defined can distinguish between an initial coherent superposition and an initial incoherent mixture of localised ion positions. 
Utilizing the properties of the trace operation, we have
\begin{equation}
{\mathrm{Tr}} \left[ \hat{\Pi}^*\hat{\rho} \right]={\mathrm{Tr}}_{\phi_{n\neq i_1,i_2}} \left[  \langle\Psi^{\ast}|\hat{\rho} |\Psi^{\ast}\rangle\right]=  \langle\Psi^{\ast}| \left[ {\mathrm{Tr}}_{\phi_{n\neq i_1,i_2}}\hat{\rho} \right] |\Psi^{\ast}\rangle,
\end{equation}
with ${\mathrm{Tr}}_{\phi_{n\neq i_1,i_2}}[\hat{\rho}_\mathrm{pure}(t)]=\hat{\rho}_\mathrm{Ir}(t)$, where $\hat{\rho}_\mathrm{Ir}(t)$ is given by Eq.~\eqref{rhored}, while  for initially mixed ion states, ${\mathrm{Tr}}_{\phi_{n\neq i_1,i_2}}[\hat{\rho}_\mathrm{mix} (t)]=\hat{\rho}^{i_1,i_1}_\mathrm{Ir}(t) +\hat{\rho}^{i_2,i_2}_\mathrm{Ir}(t)$. (See Appendix~\ref{app:localrho} for  more details concerning the reduced density matrix $\hat{\rho}_\mathrm{Ir}$.) Using the equalities $\langle\Psi^{\ast}| \rho^{i_1,i_1}_\mathrm{Ir}|\Psi^{\ast}\rangle=\langle\Psi^{\ast}| \rho^{i_2,i_2}_\mathrm{Ir}|\Psi^{\ast}\rangle$ and $|\langle\Psi^{\ast}| \rho^{i_1,i_2}_\mathrm{Ir}|\Psi^{\ast}\rangle|=|\langle\Psi^{\ast}| \rho^{i_2,i_1}_\mathrm{Ir}|\Psi^{\ast}\rangle|$, the ion-ruler coherence \eqref{ratio1} then simplifies to
\begin{equation}
C^*=\frac{|\langle\Psi^{\ast}| \hat{\rho}^{i_1,i_2}_\mathrm{Ir}|\Psi^{\ast}\rangle|}{\langle\Psi^{\ast}| \hat{\rho}^{i_1,i_1}_\mathrm{Ir}|\Psi^{\ast}\rangle}.
\label{ratio2}
\end{equation}
%\begin{figure}
%  \centering
%\includegraphics[width=.7\linewidth]{Cstarplot.pdf}
%\caption{Joint ion-ruler measurement coherence $C^*$ versus ion site separation $|i_1-i_2|$ for ruler dipole number $N=41$. The other parameters used are $c=0.1$, $\lambda=2$, and $k_r=m_r=N$, and $\hbar=1$.}
%\label{cstarplot}
%\end{figure}
\begin{figure}[htbp]
    \centering
    \includegraphics[width=0.6\linewidth]{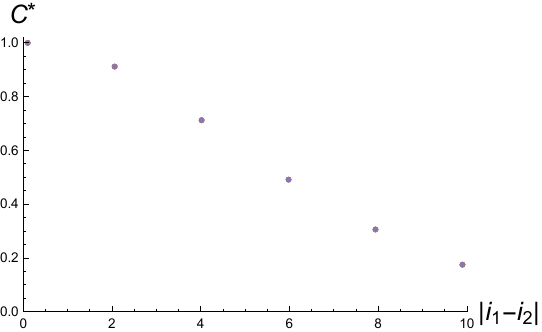}
    \caption{Joint ion-ruler measurement coherence $C^*$ versus ion site separation $|i_1-i_2|$ for ruler dipole number $N=41$. The other parameters used are $c=0.1$, $\lambda=2$, and $k_r=m_r=N$, and $\hbar=1$.}
    \label{cstarplot}
\end{figure}

Figure~\ref{cstarplot} compares the joint ion-ruler  measurement coherence $C^*$ dependence on the ion-superposition separation $|i_1-i_2|$ for ion-ruler coupling strength $\lambda=2$ and example ruler length $N=41$ (i.e., the same parameters as considered above in Fig. \ref{average2a}). The precision parameter is chosen to be $c=0.1$; smaller values of $c$ do not lead to any significant increases in $C^*$. 
The ion-ruler coherence $C^*$ decreases the further apart the localized states are in the superposition. If we were to choose a larger ion-ruler coupling, e.g., $\lambda=25$, the  coherence $C^*$ immediately drops to a negligible value: for $|i_1-i_2|=2$, we have $C^*\approx 10^{-6}$.  Thus, the ruler must operate in the weak measurement regime in order to be able to distinguish between superpositions and mixtures of localized  ion position states for a range of superposition separations.   
The qualitative trend of decreasing ion-ruler joint  coherence with increasing coupling strength is related to the nonlocal entanglement that develops between the ion and all of the ruler dipoles when they interact---a consequence of the one-dimensional ruler model; partially tracing out the other ruler dipoles $n\neq i_1,i_2$ results in decoherence. 
For a more realistic two or three dimensional mass-spring model of an extended material ruler, we expect that the ion-ruler entanglement will be more localised to the dipoles in the neighbourhood of the localised ion positions, resulting in a weaker decrease in ion-ruler coherence $C^*$  with increasing superposition separation $|i_1-i_2|$ (and perhaps allowing for the quantum ruler to operate in the strong measurement regime with comparatively larger ion-ruler coupling strengths).  Another possible way to increase the ion-ruler  coherence $C^*$ would be to include more dipole sites in the ion-ruler joint measurement projector construction, although at the expense of having a less accurate measure of the ion's location.

\section{Conclusions}

In this work, we have introduced a concrete model of a quantum position measurement device: a quantum ruler, which interacts with an ion, whose position we would like to measure. We then constructed relational observables on the joint system composed of the ruler and the ion, and showed that such a measurement procedure can distinguish between the cases in which the ion is prepared in a mixed state or in a quantum superposition state in the position basis. This generalises the usual position measurement, which localises the measured system around a single position.

This work constitutes the first step towards the long-term goal of bridging the gap between the abstract relational observables defined in quantum gravity approaches and physical quantities that can be measured in the laboratory with concrete operational procedures. Here, we have considered the simplest case, by restricting ourselves to i)~a non-relativistic and static measured system and ii)~the simplest possible measurement, acting non-trivially only on two sites of the ruler at once. 

In the future, it will be important to extend this approach. For what concerns the measured system, one could allow for some non-trivial dynamics of the ion, such as uniform velocity or acceleration. In these scenarios, one might study the emission of, respectively, Cherenkov~\cite{Cherenkov:1934ilx} or Unruh~\cite{unruh1976notes} radiation from the ion, and the measurement procedure should be adapted to capture the properties of the radiation. 

The quantum ruler model we introduce here can be easily generalised to quantum field theory, by taking the continuum limit in the distance between the sites. From a fundamental perspective, a field-theoretic description of the measurement apparatus is desirable to relate more directly the operational results of this work to quantum gravity approaches.

Another direction is to refine the measurement model to involve a larger number of sites of the ruler, either as an extended two or three-dimensional lattice. We speculate that this could increase the difference between the response of the ruler when the ion is in a mixed state versus in a quantum superposition state. The reason is that the ruler responds to the interaction with the ion more locally, with the nearest-neighbour induced dipole distortions becoming more diluted as we move radially away from the localised ion position.

Conceptually, a motivation for introducing the model of the quantum ruler is previous work on quantum reference frames~\cite{dewitt1967quantum, aharonov1, rovelli_quantum, brs_review, katz2015mesoscopic, Giacomini:2017zju}, namely reference frames associated to physical systems, which can be in a quantum superposition or entangled relative to each other. One general goal of the quantum reference frames programme is to substitute 
the abstract description of a coordinate system by providing a more physical one, which relies on measurements performed on physical objects. It would be interesting to associate a quantum reference frame to a quantum ruler, and compare the perspectives of different quantum rulers. On the one hand, this would be an important step to achieve a relational perspective on nonclassical spacetime closer to research in quantum gravity, where quantum reference frames are extended material systems. On the other hand, it would also identify a procedure to measure the position of a quantum system relative to a quantum reference frame, whose concrete implementation is still an open question in the field.

\acknowledgements{We thank S.A. Ahmad and A.R.H. Smith for very helpful discussions. F.N. is supported in part by: Nippon Telegraph and Telephone Corporation (NTT) Research, the Japan Science and Technology Agency (JST) [via the Quantum Leap Flagship Program (Q-LEAP), and the Moonshot R$\&$D Grant Number JPMJMS2061], the Asian Office of Aerospace Research and Development (AOARD) (via Grant No. FA2386-20-1-4069), and the Office of Naval Research (ONR) (via Grant No. N62909-23-1-2074). F.G. acknowledges support from Perimeter Institute for Theoretical Physics. Research at Perimeter Institute is supported in part by the Government of Canada through the Department of Innovation, Science and Economic Development and by the Province of Ontario through the Ministry of Colleges and Universities. F.G. acknowledges support from the Swiss National Science Foundation via the Ambizione Grant PZ00P2-208885, from the ETH Zurich Quantum Center, and from the ID\#~62312 grant from the John Templeton Foundation, as part of the \href{https://www.templeton.org/grant/the-quantuminformation-structure-ofspacetime-qiss-second-phase}{‘The Quantum Information Structure of Spacetime, Second Phase (QISS 2)’ Project}. The opinions expressed in this publication are those of the authors and do not necessarily reflect the views of the John Templeton Foundation. M.P.B. and H.W. acknowledge support from the U.S.  National Science Foundation  under grant number PHY-2011382.}
\bibliographystyle{quantum}
\bibliography{biblioRuler}

\begin{thebibliography}{10}

\bibitem{Peres:2002ip}
Asher Peres, Petra~F. Scudo, and Daniel~R. Terno.
\newblock ``{Quantum entropy and special relativity}''.
\newblock \href{https://dx.doi.org/10.1103/PhysRevLett.88.230402}{Phys. Rev.
  Lett. {\bf 88}, 230402}~(2002).
\newblock
  \href{http://arxiv.org/abs/quant-ph/0203033}{arXiv:quant-ph/0203033}.

\bibitem{bauke2014relativistic}
Heiko Bauke, Sven Ahrens, Christoph~H. Keitel, and Rainer Grobe.
\newblock ``What is the relativistic spin operator?''.
\newblock \href{https://dx.doi.org/10.1088/1367-2630/16/4/043012}{New Journal
  of Physics {\bf 16}, 043012}~(2014).
\newblock  \href{http://arxiv.org/abs/1303.3862}{arXiv:1303.3862}.

\bibitem{erker2017autonomous}
Paul Erker, Mark~T Mitchison, Ralph Silva, Mischa~P Woods, Nicolas Brunner, and
  Marcus Huber.
\newblock ``Autonomous quantum clocks: does thermodynamics limit our ability to
  measure time?''.
\newblock \href{https://dx.doi.org/10.1103/PhysRevX.7.031022}{Phys. Rev. X {\bf
  7}, 031022}~(2017).
\newblock  \href{http://arxiv.org/abs/1609.06704}{arXiv:1609.06704}.

\bibitem{woods2018quantum}
Mischa~P Woods, Ralph Silva, Gilles P{\"u}tz, Sandra Stupar, and Renato Renner.
\newblock ``Quantum clocks are more accurate than classical ones''.
\newblock \href{https://dx.doi.org/10.1103/PRXQuantum.3.010319}{PRX Quantum
  {\bf 3}, 010319}~(2022).
\newblock  \href{http://arxiv.org/abs/1806.00491}{arXiv:1806.00491}.

\bibitem{Tino:2007az}
Guglielmo~M. Tino et~al.
\newblock ``{Atom interferometers and optical atomic clocks: New quantum
  sensors for fundamental physics experiments in space}''.
\newblock \href{https://dx.doi.org/10.1016/j.nuclphysbps.2006.12.061}{Nucl.
  Phys. B Proc. Suppl. {\bf 166}, 159--165}~(2007).

\bibitem{zych_interferometric}
Magdalena Zych, Fabio Costa, Igor Pikovski, and {\v{C}}aslav Brukner.
\newblock ``Quantum interferometric visibility as a witness of general
  relativistic proper time''.
\newblock \href{https://dx.doi.org/10.1038/ncomms1498}{Nature communications
  {\bf 2}, 505}~(2011).
\newblock  \href{http://arxiv.org/abs/1105.4531}{arXiv:1105.4531}.

\bibitem{Giulini:2011xf}
Domenico Giulini.
\newblock ``Equivalence principle, quantum mechanics, and atom-interferometric
  tests''.
\newblock \href{https://dx.doi.org/10.1007/978-3-0348-0043-3_16}{Pages
  345--370}.
\newblock Springer Basel. Basel~(2012).
\newblock  \href{http://arxiv.org/abs/1105.0749}{arXiv:1105.0749}.

\bibitem{zych}
Magdalena Zych, Igor Pikovski, Fabio Costa, and {\v{C}}aslav Brukner.
\newblock ``General relativistic effects in quantum interference of
  “clocks”''.
\newblock \href{https://dx.doi.org/10.1088/1742-6596/723/1/012044}{Journal of
  Physics: Conference Series {\bf 723}, 012044}~(2016).
\newblock  \href{http://arxiv.org/abs/1607.04022}{arXiv:1607.04022}.

\bibitem{Roura:2018cfg}
Albert Roura.
\newblock ``{Gravitational redshift in quantum-clock interferometry}''.
\newblock \href{https://dx.doi.org/10.1103/PhysRevX.10.021014}{Phys. Rev. X
  {\bf 10}, 021014}~(2020).
\newblock  \href{http://arxiv.org/abs/1810.06744}{arXiv:1810.06744}.

\bibitem{Knight:1996sxc}
Peter Knight.
\newblock ``Measuring quantum states with quantum rulers''.
\newblock In International Quantum Electronics Conference.
\newblock Page FF2.
\newblock Optica Publishing Group~(1996).
\newblock
  url:~\href{https://opg.optica.org/abstract.cfm?URI=IQEC-1996-FF2}{opg.optica.org/abstract.cfm?URI=IQEC-1996-FF2}.

\bibitem{ralph2002coherent}
Timothy~C. Ralph.
\newblock ``Coherent superposition states as quantum rulers''.
\newblock \href{https://dx.doi.org/10.1103/PhysRevA.65.042313}{Phys. Rev. A
  {\bf 65}, 042313}~(2002).
\newblock
  \href{http://arxiv.org/abs/quant-ph/0109106}{arXiv:quant-ph/0109106}.

\bibitem{dewitt1967quantum}
Bryce~S. DeWitt.
\newblock ``{Quantum theory of gravity. I. The canonical theory}''.
\newblock \href{https://dx.doi.org/10.1103/PhysRev.160.1113}{Phys. Rev. {\bf
  160}, 1113}~(1967).

\bibitem{Kuchar:1990vy}
Karel~V. Kuchar and Charles~G. Torre.
\newblock ``{Gaussian reference fluid and interpretation of quantum
  geometrodynamics}''.
\newblock \href{https://dx.doi.org/10.1103/PhysRevD.43.419}{Phys. Rev. D {\bf
  43}, 419--441}~(1991).

\bibitem{Brown:1994py}
J.~David Brown and Karel~V. Kuchar.
\newblock ``{Dust as a standard of space and time in canonical quantum
  gravity}''.
\newblock \href{https://dx.doi.org/10.1103/PhysRevD.51.5600}{Phys. Rev. D {\bf
  51}, 5600--5629}~(1995).
\newblock  \href{http://arxiv.org/abs/gr-qc/9409001}{arXiv:gr-qc/9409001}.

\bibitem{Brown:1995fj}
J.~David Brown and Donald Marolf.
\newblock ``{On relativistic material reference systems}''.
\newblock \href{https://dx.doi.org/10.1103/PhysRevD.53.1835}{Phys. Rev. D {\bf
  53}, 1835--1844}~(1996).
\newblock  \href{http://arxiv.org/abs/gr-qc/9509026}{arXiv:gr-qc/9509026}.

\bibitem{butterfield2001spacetime}
George F.~R. Ellis and Rituparno Goswami.
\newblock ``{Space time and the passage of time}''.
\newblock In Abhay Ashtekar and Vesselin Petkov, editors, {Springer Handbook of
  Spacetime}.
\newblock \href{https://dx.doi.org/10.1007/978-3-642-41992-8_13}{Pages
  243--264}.
\newblock Springer~(2014).
\newblock  \href{http://arxiv.org/abs/1208.2611}{arXiv:1208.2611}.

\bibitem{Rovelli:2004tv}
Carlo Rovelli.
\newblock ``Quantum gravity''.
\newblock
  \href{https://dx.doi.org/https://doi.org/10.1017/CBO9780511755804}{Cambridge
  Monographs on Mathematical Physics}. Cambridge University Press. ~(2004).

\bibitem{rovelli_relational}
Carlo Rovelli.
\newblock ``Relational quantum mechanics''.
\newblock Int. J. Theor. Phys. {\bf 35}, 1637--1678~(1996).
\newblock
  \href{http://arxiv.org/abs/quant-ph/9609002}{arXiv:quant-ph/9609002}.

\bibitem{anderson2017problem}
Edward Anderson.
\newblock ``The problem of time''.
\newblock
  \href{https://dx.doi.org/https://doi.org/10.1007/978-3-319-58848-3}{Springer}.
  ~(2017).

\bibitem{cirac1995}
Juan~I Cirac and Peter Zoller.
\newblock ``Quantum computations with cold trapped ions''.
\newblock \href{https://dx.doi.org/10.1103/PhysRevLett.74.4091}{Phys. Rev.
  Lett. {\bf 74}, 4091--4094}~(1995).

\bibitem{chen2023}
Wentao Chen, Yao Lu, Shuaining Zhang, Kuan Zhang, Guanhao Huang, Mu~Qiao,
  Xiaolu Su, Jialiang Zhang, Jing-Ning Zhang, Leonardo Banchi, M.~S. Kim, and
  Kihwan Kim.
\newblock ``Scalable and programmable phononic network with trapped ions''.
\newblock \href{https://dx.doi.org/10.1038/s41567-023-01952-5}{Nat. Phys.Pages
  1--7}~(2023).
\newblock  \href{http://arxiv.org/abs/2207.06115}{arXiv:2207.06115}.

\bibitem{vetsch2010}
E.~Vetsch, D.~Reitz, G.~Sagu\'e, R.~Schmidt, S.~T. Dawkins, and
  A.~Rauschenbeutel.
\newblock ``Optical interface created by laser-cooled atoms trapped in the
  evanescent field surrounding an optical nanofiber''.
\newblock \href{https://dx.doi.org/10.1103/PhysRevLett.104.203603}{Phys. Rev.
  Lett. {\bf 104}, 203603}~(2010).
\newblock  \href{http://arxiv.org/abs/0912.1179}{arXiv:0912.1179}.

\bibitem{bruschi2019time}
David~Edward Bruschi.
\newblock ``Time evolution of coupled multimode and multiresonator
  optomechanical systems''.
\newblock \href{https://dx.doi.org/10.1063/1.5106409}{J. Math. Phys. {\bf 60},
  062105}~(2019).
\newblock  \href{http://arxiv.org/abs/1812.06879}{arXiv:1812.06879}.

\bibitem{zurek1991}
Wojciech~H. Zurek.
\newblock ``{Decoherence and the Transition from Quantum to Classical}''.
\newblock \href{https://dx.doi.org/10.1063/1.881293}{Physics Today {\bf 44},
  36--44}~(1991).

\bibitem{Lynden-Bell:1995cmj}
Donald Lynden-Bell and Joseph Katz.
\newblock ``{Classical mechanics without absolute space}''.
\newblock \href{https://dx.doi.org/10.1103/PhysRevD.52.7322}{Phys. Rev. D {\bf
  52}, 7322--7324}~(1995).
\newblock
  \href{http://arxiv.org/abs/astro-ph/9509158}{arXiv:astro-ph/9509158}.

\bibitem{Dittrich:2004cb}
Bianca Dittrich.
\newblock ``{Partial and complete observables for Hamiltonian constrained
  systems}''.
\newblock \href{https://dx.doi.org/10.1007/s10714-007-0495-2}{Gen. Rel. Grav.
  {\bf 39}, 1891--1927}~(2007).
\newblock  \href{http://arxiv.org/abs/gr-qc/0411013}{arXiv:gr-qc/0411013}.

\bibitem{Tambornino:2011vg}
Johannes Tambornino.
\newblock ``{Relational Observables in Gravity: a Review}''.
\newblock \href{https://dx.doi.org/10.3842/SIGMA.2012.017}{SIGMA {\bf 8},
  017}~(2012).
\newblock  \href{http://arxiv.org/abs/1109.0740}{arXiv:1109.0740}.

\bibitem{Hoehn:2019fsy}
Philipp~A. Hoehn, Alexander R.~H. Smith, and Maximilian P.~E. Lock.
\newblock ``{Trinity of relational quantum dynamics}''.
\newblock \href{https://dx.doi.org/10.1103/PhysRevD.104.066001}{Phys. Rev. D
  {\bf 104}, 066001}~(2021).
\newblock  \href{http://arxiv.org/abs/1912.00033}{arXiv:1912.00033}.

\bibitem{Giesel:2020xnb}
Kristina Giesel, Bao-Fei Li, and Parampreet Singh.
\newblock ``{Relating dust reference models to conventional systems in
  manifestly gauge invariant perturbation theory}''.
\newblock \href{https://dx.doi.org/10.1103/PhysRevD.104.023501}{Phys. Rev. D
  {\bf 104}, 023501}~(2021).
\newblock  \href{http://arxiv.org/abs/2012.14443}{arXiv:2012.14443}.

\bibitem{Bojowald:2021uqo}
Martin Bojowald, Luiz Martinez, and Garrett Wendel.
\newblock ``{Relational evolution with oscillating clocks}''.
\newblock \href{https://dx.doi.org/10.1103/PhysRevD.105.106020}{Phys. Rev. D
  {\bf 105}, 106020}~(2022).
\newblock  \href{http://arxiv.org/abs/2110.07702}{arXiv:2110.07702}.

\bibitem{Baldazzi:2021fye}
Alessio Baldazzi, Kevin Falls, and Renata Ferrero.
\newblock ``{Relational observables in asymptotically safe gravity}''.
\newblock \href{https://dx.doi.org/10.1016/j.aop.2022.168822}{Annals Phys. {\bf
  440}, 168822}~(2022).
\newblock  \href{http://arxiv.org/abs/2112.02118}{arXiv:2112.02118}.

\bibitem{kofler2007classical}
Johannes Kofler and \ifmmode \check{C}\else~\v{C}\fi{}aslav Brukner.
\newblock ``Classical world arising out of quantum physics under the
  restriction of coarse-grained measurements''.
\newblock \href{https://dx.doi.org/10.1103/PhysRevLett.99.180403}{Phys. Rev.
  Lett. {\bf 99}, 180403}~(2007).
\newblock
  \href{http://arxiv.org/abs/quant-ph/0609079}{arXiv:quant-ph/0609079}.

\bibitem{dakic2016classical}
Borivoje Daki\'c and {\v{C}}aslav Brukner.
\newblock ``{The Classical Limit of a Physical Theory and the Dimensionality of
  Space}''.
\newblock \href{https://dx.doi.org/10.1007/978-94-017-7303-4_8}{Fundam. Theor.
  Phys. {\bf 181}, 249--282}~(2016).
\newblock  \href{http://arxiv.org/abs/1307.3984}{arXiv:1307.3984}.

\bibitem{oreshkov2005weak}
Ognyan Oreshkov and Todd~A. Brun.
\newblock ``Weak measurements are universal''.
\newblock \href{https://dx.doi.org/10.1103/PhysRevLett.95.110409}{Phys. Rev.
  Lett. {\bf 95}, 110409}~(2005).
\newblock
  \href{http://arxiv.org/abs/quant-ph/0503017}{arXiv:quant-ph/0503017}.

\bibitem{schmid1983}
Albert Schmid.
\newblock ``Diffusion and localization in a dissipative quantum system''.
\newblock \href{https://dx.doi.org/10.1103/PhysRevLett.51.1506}{Phys. Rev.
  Lett. {\bf 51}, 1506--1509}~(1983).

\bibitem{fisher1985}
Matthew P.~A. Fisher and Wilhelm Zwerger.
\newblock ``Quantum brownian motion in a periodic potential''.
\newblock \href{https://dx.doi.org/10.1103/PhysRevB.32.6190}{Phys. Rev. B {\bf
  32}, 6190--6206}~(1985).

\bibitem{aslangul1985}
C.~Aslangul, N.~Pottier, and D.~Saint-James.
\newblock ``Quantum ohmic dissipation: Particle on a one-dimensional periodic
  lattice''.
\newblock
  \href{https://dx.doi.org/https://doi.org/10.1016/0375-9601(85)90570-5}{Physics
  Letters A {\bf 111}, 175--178}~(1985).

\bibitem{baumgratz2014quantifying}
Tillmann Baumgratz, Marcus Cramer, and Martin~B Plenio.
\newblock ``Quantifying coherence''.
\newblock \href{https://dx.doi.org/10.1103/PhysRevLett.113.140401}{Phys. Rev.
  Lett. {\bf 113}, 140401}~(2014).
\newblock  \href{http://arxiv.org/abs/1311.0275}{arXiv:1311.0275}.

\bibitem{Cherenkov:1934ilx}
Pavel~A. Cherenkov.
\newblock ``{Visible luminescence of pure liquids under the influence of
  \ensuremath{\gamma}-radiation}''.
\newblock \href{https://dx.doi.org/10.3367/UFNr.0093.196710n.0385}{Dokl. Akad.
  Nauk SSSR {\bf 2}, 451--454}~(1934).

\bibitem{unruh1976notes}
William~G. Unruh.
\newblock ``Notes on black-hole evaporation''.
\newblock \href{https://dx.doi.org/10.1103/PhysRevD.14.870}{Phys. Rev. D {\bf
  14}, 870--892}~(1976).

\bibitem{aharonov1}
Yakir Aharonov and Leonard Susskind.
\newblock ``{Charge Superselection Rule}''.
\newblock \href{https://dx.doi.org/10.1103/PhysRev.155.1428}{Phys. Rev. {\bf
  155}, 1428--1431}~(1967).

\bibitem{rovelli_quantum}
Carlo Rovelli.
\newblock ``Relational quantum mechanics''.
\newblock
  \href{https://dx.doi.org/https://doi.org/10.1007/BF02302261}{International
  journal of theoretical physics {\bf 35}, 1637--1678}~(1996).

\bibitem{brs_review}
Stephen~D. Bartlett, Terry Rudolph, and Robert~W. Spekkens.
\newblock ``Reference frames, superselection rules, and quantum information''.
\newblock \href{https://dx.doi.org/10.1103/RevModPhys.79.555}{Rev. Mod. Phys.
  {\bf 79}, 555--609}~(2007).
\newblock
  \href{http://arxiv.org/abs/quant-ph/0610030}{arXiv:quant-ph/0610030}.

\bibitem{katz2015mesoscopic}
B.~N. Katz, M.~P. Blencowe, and K.~C. Schwab.
\newblock ``Mesoscopic mechanical resonators as quantum noninertial reference
  frames''.
\newblock \href{https://dx.doi.org/10.1103/PhysRevA.92.042104}{Phys. Rev. A
  {\bf 92}, 042104}~(2015).
\newblock  \href{http://arxiv.org/abs/1409.2137}{arXiv:1409.2137}.

\bibitem{Giacomini:2017zju}
Flaminia Giacomini, Esteban Castro-Ruiz, and {\v C}aslav Brukner.
\newblock ``Quantum mechanics and the covariance of physical laws in quantum
  reference frames''.
\newblock \href{https://dx.doi.org/10.1038/s41467-018-08155-0}{Nat. Commun.
  {\bf 10}, 494}~(2019).
\newblock  \href{http://arxiv.org/abs/1712.07207}{arXiv:1712.07207}.

\end{thebibliography}
\appendix

\section{Lagrangian formulation of the ion-ruler system}
\label{app:Lagrangian}
The individual Lagrangians of the ion and the ruler subsystems are
\begin{eqnarray}
L_I &=& \frac{1}{2} M_I \dot{x}_I^2 , 
\label{ldetector1}
\end{eqnarray}

\begin{equation}
L_r =  \frac{1}{2} m_r \sum_{n=-\frac{N-1}{2}}^{\frac{N-1}{2}} \dot{x}_{r,n}^2 - \frac{1}{2} k_r \sum_{n=-\frac{N-1}{2}}^{\frac{N-3}{2}} (x_{r,n+1} - x_{r,n} - a_r)^2. 
\label{ruler1}
\end{equation}
Introducing the ruler centre-of-mass and relative coordinates, we have
\begin{equation}
x_\mathrm{r CM} = \frac{1} {N} \sum_{n=-\frac{N-1}{2}} ^{\frac{N-1}{2}} x_{r,n},
\end{equation}
\begin{equation}
\phi_{n} = x_{r,n} - x_\mathrm{r CM}-n a_r.
\end{equation}
The coordinate $\phi_{n}$ gives the displacement of the $n$th dipole relative to its classical equilibrium position. While a more natural ruler coordinate would be $\tilde{x}_{r,n}=x_{r,n}-x_{\mathrm{r CM}}$, marking the distance from the ruler's midpoint in equilibrium, the former $\phi_{n}$ coordinates are more suited for indicating local  elastic displacements induced by the nearby ion; we can always easily convert to a ruler length coordinate by considering $\phi_n+n a_r=\tilde{x}_{r,n}$.  
In terms of the above coordinates, the ruler Lagrangian becomes
\begin{equation}
L_r =  \frac{1}{2} M_r \dot{x}_{\mathrm{r CM}}^2+\frac{1}{2} m_r \sum_{n=-\frac{N-1}{2}}^{\frac{N-1}{2}} \dot{\phi}_{n}^2 - \frac{1}{2} k_r \sum_{n=-\frac{N-1}{2}}^{\frac{N-3}{2}} (\phi_{n+1} - \phi_{n})^2, 
\label{ruler2}
\end{equation}
where $M_r=N m_r$ is the  ruler's total mass. The Coulomb interaction potential energy between the ruler and ion is
\begin{eqnarray}
V_{Ir} &=& -\frac{q_I q} {4\pi \epsilon_0} \sum_{n=-\frac{N-1}{2}}^{\frac{N-1}{2}} \left[\frac{1} {\sqrt{(w-l/2)^2+(x_I-x_{r,n})^2}} - \frac{1} {\sqrt{(w+l/2)^2+(x_I-x_{r,n})^2}}\right] \cr
&\approx& -\frac{q_I {\mathsf{p}}_r w} {4\pi \epsilon_0} \sum_{n=-\frac{N-1}{2}}^{\frac{N-1}{2}} \left[w^2+(x_I-x_{r,n})^2\right]^{-3/2},
\label{coulombeq}
\end{eqnarray}
where the approximation is valid under the limit $l\ll w$, with $l$ the distance between the ruler dipole charges $q$ and $-q$, ${\mathsf{p}}_r=q l$ is the ruler atom electric dipole moment, and $x_{r,n} = \phi_n + na_r+x_{\mathrm{r CM}}$. The interaction potential is further approximated as
\begin{equation}
  {V}_{I r}  \approx \left.-\frac{q_I {\mathsf{p}}_r w} {4\pi \epsilon_0} \sum_{n=-\frac{N-1}{2}} ^{\frac{N-1}{2}}\left[({x}_{In}^{2} + w^2)^{-3/2} + 3 {x}_{In} {\phi}_n  ({x}_{In}^{2} + w^2)^{-5/2} \right]\right|_{x_{In}=x_I - na_r - x_{\mathrm{r CM}}},
  \label{ion-dipoleVeq}
\end{equation}
under the limit $\phi_n\ll w$.

\section{The equivalence of constraints}
\label{app:constraints}
 Starting with the ruler Lagrangian of Eq.\,(\ref{ruler2})
\begin{equation}
L_r =  \frac{1}{2} M_r \dot{x}_{\mathrm{r CM}}^2+\frac{1}{2} m_r \sum_{n=-\frac{N-1}{2}}^{\frac{N-1}{2}} \dot{\phi}_{n}^2 - \frac{1}{2} k_r \sum_{n=-\frac{N-1}{2}}^{\frac{N-3}{2}} (\phi_{n+1} - \phi_{n})^2, 
\end{equation}
and imposing the constraint of Eq.\,(\ref{constrainteq1})
\begin{equation}
    x_{\mathrm{rCM}}=\frac{1}{N}\sum_{n=-\frac{N-1}{2}}^{\frac{N-1}{2}}x_{r,n}=0,
\end{equation}
we obtain the corresponding quantized ruler Hamiltonian~(\ref{quantizedruler}). 

On the other hand, starting with Hamiltonian (\ref{quantizedruler}) and substituting in
\begin{equation}
    \hat{a}_{\alpha}= \sqrt{\frac{m_r\Omega_{\alpha}}{2\hbar}} \hat{x}_{\alpha}+\frac{i}{\sqrt{2 \hbar m_r\Omega_{\alpha}}} \hat{p}_{\alpha},
\end{equation}
we obtain the ruler Hamiltonian in terms of the nonlocal canonical coordinate pairs $\hat{x}_{\alpha}$ and $\hat{p}_{\alpha}$:
\begin{equation}
 \hat{H}_r = \sum_{\alpha=1}^{N-1} \frac{\hat{p}_{\alpha}^2}{2m_r} + \frac{1}{2} m_r \sum_{\alpha=1}^{N-1} \Omega_\alpha^2 \hat{x}_{\alpha}^2, 
\end{equation}
and the corresponding Lagrangian is
\begin{equation}
 {L}_r = \sum_{\alpha=1}^{N-1} \frac{m_r \dot{{x}}_{\alpha}^2}{2} - \frac{1}{2} m_r \sum_{\alpha=1}^{N-1} \Omega_\alpha^2 {x}_{\alpha}^2. 
\end{equation}
Using the relation $x_\alpha=\sum_{n=-\frac{N-1}{2}} ^{\frac{N-1}{2}}  u_{\alpha,n}\phi_n$ to express the Lagrangian in terms of the local coordinates $\phi_n$, we obtain
\begin{eqnarray}
 {L}_r &=& \sum_{\alpha=1}^{N-1} \left(\frac{1}{2} m_r\sum_{n,n^\prime=-\frac{N-1}{2}}^{\frac{N-1}{2}} u_{\alpha,n} u_{\alpha,n^\prime} \dot{{\phi}}_n \dot{{\phi}}_{n^\prime} - \frac{1}{2} m_r  \Omega_\alpha^2 \sum_{n,n^\prime=-\frac{N-1}{2}}^{\frac{N-1}{2}} u_{\alpha,n} u_{\alpha,n^\prime} {\phi}_n {\phi}_n^\prime \right)\cr
 &=& \sum_{n,n^\prime=-\frac{N-1}{2}}^{\frac{N-1}{2}}\left(\frac{1}{2} m_r \dot{{\phi}}_n \dot{{\phi}}_{n^\prime} \sum_{\alpha=1}^{N-1} u_{\alpha,n} u_{\alpha,n^\prime} - \frac{1}{2} m_r {\phi}_n {\phi}_{n^\prime} \sum_{\alpha=1}^{N-1} \Omega_\alpha^2 u_{\alpha,n} u_{\alpha,n^\prime} \right)\cr
 &=& \sum_{n=-\frac{N-1}{2}}^{\frac{N-1}{2}} \frac{1}{2} m_r \dot{{\phi}}_n^2-\frac{m_r}{N}\left( \sum_{n=-\frac{N-1}{2}}^{\frac{N-1}{2}} \dot{{\phi}}_n \right)^2 - \frac{1}{2} k_r \sum_{n=-\frac{N-1}{2}}^{\frac{N-3}{2}} ({\phi}_{n+1} - {\phi}_{n})^2\cr
 &=& \sum_{n=-\frac{N-1}{2}}^{\frac{N-1}{2}} \frac{1}{2} m_r \dot{{\phi}}_n^2 - \frac{1}{2} k_r \sum_{n=-\frac{N-1}{2}}^{\frac{N-3}{2}} ({\phi}_{n+1} - {\phi_{n}})^2.
\end{eqnarray}
The last equality holds due to the constraint on the relative coordinates [Eq.~\eqref{constrainteq2}]:
\begin{equation}
 \sum_{n=-\frac{N-1}{2}}^{\frac{N-1}{2}} \hat{\phi}_n=0.
\end{equation}
We then recover the ruler Lagrangian~(\ref{ruler2}) when the constraint $x_{\mathrm{rCM}}=0$ is satisfied. Therefore, the constraint $\sum_{n=-\frac{N-1}{2}}^{\frac{N-1}{2}} {\phi}_n=0$ is effectively the same as the constraint $x_{\mathrm{rCM}}=0$.

\section{Free ruler dynamics}
\label{app:freeruler}
The free ruler's dipole displacement equations of motion can be easily solved in terms of normal modes as we now show. From the Lagrangian (\ref{ruler2}), 
the equations of motion are
\begin{equation}
    \ddot{\phi}_{n}=\omega_r^2\left(\phi_{n-1}-2 \phi_n+\phi_{n+1}\right),\, -\frac{1}{2}(N-3)\leq n\leq +\frac{1}{2}(N-3),
    \label{eom}
\end{equation}
with boundary conditions
\begin{eqnarray}
\ddot{\phi}_{\frac{N-1}{2}}&=&-\omega_r^2 \left(\phi_{\frac{N-1}{2}}- \phi_{\frac{N-3}{2}}\right)\label{bc1}\\
\ddot{\phi}_{-\frac{N-1}{2}}&=&-\omega_r^2 \left(\phi_{-\frac{N-1}{2}}- \phi_{-\frac{N-3}{2}}\right),
\end{eqnarray}
where $\omega_r=\sqrt{k_r/m_r}$.
It is convenient to introduce symmetric and antisymmetric coordinates:
\begin{equation}
    \phi^s_n=\frac{1}{2}\left(\phi_n+\phi_{-n}\right);\, \phi^a_n=\frac{1}{2}\left(\phi_n-\phi_{-n}\right), 
\end{equation}
where $\phi^s$ and $\phi^a$ still satisfy the equation of motion (\ref{eom}) and the boundary condition (\ref{bc1}), but where now we have the following boundary conditions at $n=0$:
\begin{eqnarray}
    &&\phi_{0}^a=0,\label{abc}\\ &&\ddot{\phi}_{0}^s=-2\omega_r^2\left(\phi_0^s-\phi_1^s\right).\label{sbc}
\end{eqnarray}
The constraint (\ref{constrainteq2}) is now imposed only on the symmetric coordinate:
\begin{equation}
    \phi_0^s+2\sum_{n=1}^{\frac{N-1}{2}}\phi_n^s=0.
    \label{constraint2eq}
\end{equation}
Consider a mode solution Ansatz of the form:
\begin{equation}
    \phi_n(t)=\cos\left(\Omega t+\varphi\right)\left[ A\cos\left(k n a_r\right)+B\sin\left(k n a_r\right)\right].
    \label{symsoleq}
\end{equation}
Substituting into the equation of motion (\ref{eom}), we obtain after some algebra the following dispersion relation between mode frequency $\Omega$ and wave number $k$:
\begin{equation}
    \Omega =2\omega_r \sin\left(\frac{k a_r}{2}\right).
\end{equation}
Imposing the antisymmetric coordinate boundary condition (\ref{abc}), we have $A=0$, while for the symmetric coordinate boundary condition (\ref{sbc}), we have $B=0$. Imposing the boundary condition (\ref{bc1}) at the free end of the ruler, we obtain
$k_{\alpha}=\frac{2\alpha\pi}{N a_r}$, $\alpha=0,1,2,\dots,\frac{N-1}{2}$ for the symmetric mode solutions, and $k_{\alpha}=\frac{\left(2\alpha+1\right)\pi}{N a_r}$, $\alpha=0,1,2,\dots,\frac{N-3}{2}$ for the antisymmetric mode solutions, where here $\alpha$ denotes the mode label. 

Putting everything together so far, the symmetric and antisymmetric normal mode solutions can be written as follows:
\begin{equation}
    \phi^s_{\alpha, n}(t)=A_{\alpha}\cos\left(\Omega^s_{\alpha}t+\varphi^s_{\alpha}\right)
    \cos\left(\frac{2\alpha n \pi}{N}\right),\, \alpha=1,2,\dots,\frac{N-1}{2},
    \label{smode}
\end{equation}
\begin{equation}
    \phi^a_{\alpha, n}(t)=B_{\alpha}\cos\left(\Omega^a_{\alpha}t+\varphi^a_{\alpha}\right)
    \sin\left[\frac{\left(2\alpha+1\right)n\pi}{N}\right], \,\alpha=0,1,2,\dots,\frac{N-3}{2}, 
    \label{amode}
\end{equation}
where $\Omega^s_{\alpha}=2\omega_r \sin\left(\frac{\alpha \pi}{N}\right)$ and $\Omega^a_{\alpha}=2\omega_r \sin\left[\frac{\left(\alpha+\frac{1}{2}\right) \pi}{N}\right]$.
Note that we do not include the zero frequency, $\alpha=0$ mode for the symmetric case, a consequence of the constraint (\ref{constraint2eq}); for all symmetric normal mode solutions (\ref{smode}) with $\alpha\geq 1$, one can verify that the constraint condition is satisfied. In other words, imposing the constraint removes the zero frequency, centre of mass mode.

The symmetric and antisymmetric normal mode frequencies can be combined as $\Omega_\alpha=2\omega_r\sin(\frac{\alpha\pi}{2N})$, $\alpha=1,2,\dots,N-1$. In terms of the above, derived normal mode solutions, an arbitrary ruler spatial coordinate solution can be expressed as a linear combination of the former as follows:
\begin{equation}
\phi_n(t)=\sum_{\alpha=1}^{N-1}\left[\cos\left(\Omega_{\alpha} t\right) x_{\alpha} (0)+\sin\left(\Omega_{\alpha} t\right) \frac{p_{\alpha} (0)}{m_r\Omega_{\alpha}}\right] u_{\alpha,n},
\label{gensolneq}
\end{equation}
where $\left(x_{\alpha}(0),p_{\alpha}(0)\right)$ are the initial mode $\alpha$ canonical position and momentum coordinates, and where
\begin{equation}
    u_{\alpha, n}=\sqrt{\frac{2}{N}} \cos \left[\frac{\alpha\pi}{N}\left(n+\frac{N}{2}\right)\right]
\end{equation}
are the orthonormal mode eigenfunctions.

With expression (\ref{gensolneq}), it is straightforward to quantize the ruler coordinates in the Heisenberg picture. In terms of the  $\alpha$ mode lowering operator:
\begin{equation}
    \hat{a}_{\alpha}(0)=\sqrt{\frac{m_r\Omega_{\alpha}}{2\hbar}} \hat{x}_{\alpha}(0)+\frac{i}{\sqrt{2 \hbar m_r\Omega_{\alpha}}} \hat{p}_{\alpha}(0),
\end{equation}
Eq. (\ref{gensolneq}) becomes
\begin{equation}
\hat{\phi}_n(t)=\sum_{\alpha=1}^{N-1}\sqrt{\frac{\hbar}{2 m_r\Omega_{\alpha}}}\left[\hat{a}_{\alpha}(0) e^{-i\Omega_{\alpha} t}+\hat{a}_{\alpha}^{ \dag}(0) e^{i\Omega_{\alpha} t}\right]u_{\alpha,n}.
\label{heissolneq}
\end{equation}
The momentum operator $\hat{\pi}_n(t)$ canonically conjugate to $\hat{\phi}_n(t)$ is
\begin{eqnarray}
\hat{\pi}_n(t)&=&-i \sum_{\alpha=1} ^{N-1} \sqrt{\frac{\hbar m_r\Omega_{\alpha}}{2}} \left[\hat{a}_{\alpha}(0) e^{-i\Omega_{\alpha} t}-\hat{a}_{\alpha}^{ \dag}(0) e^{i\Omega_{\alpha} t}\right] u_{\alpha,n}.
\end{eqnarray}

\section{Density matrix elements in the local basis}
\label{app:localrho}

Take one element of the reduced density matrix Eq.~(\ref{rhored}), for example
\begin{eqnarray}
\rho_\mathrm{Ir}^{i_1,i_2}(t)&=&\frac{\sqrt{N}}{2}|i_1\rangle \langle i_2| \otimes\int d\phi_{i_1} d\phi_{i_1}^{\prime}d\phi_{i_2} d\phi_{i_2}^{\prime}|\phi_{i_1}\rangle \langle \phi_{i_1}^{\prime}| \otimes |\phi_{i_2}\rangle \langle \phi_{i_2}^{\prime}| \cr
&&\times\prod_{n\neq i_1,i_2}\int d\phi_n \prod_{\alpha=1}^{N-1} \psi_{i_1} \left(\sum_n \tilde{u}_{\alpha,n} \phi_n,t\right)\psi_{i_2}^\ast \left(\sum_n \tilde{u}_{\alpha,n} \phi_n^{\prime},t \right).
\label{rhoi1i2}
\end{eqnarray}
The second line of Eq. (\ref{rhoi1i2}) is a function of $\phi_{i_1},  \phi_{i_1}^{\prime}, \phi_{i_2}$ and $\phi_{i_2}^{\prime}$. The other reduced density elements are
\begin{eqnarray}
\rho_\mathrm{Ir}^{i_1,i_1}(t)&=&\frac{\sqrt{N}}{2}|i_1\rangle \langle i_1| \otimes\int d\phi_{i_1} d\phi_{i_1}^{\prime}d\phi_{i_2} d\phi_{i_2}^{\prime}|\phi_{i_1}\rangle \langle \phi_{i_1}^{\prime}| \otimes |\phi_{i_2}\rangle \langle \phi_{i_2}^{\prime}| \cr
&&\times \prod_{n\neq i_1,i_2}\int d\phi_n \prod_{\alpha=1}^{N-1} \psi_{i_1} \left(\sum_n \tilde{u}_{\alpha,n} \phi_n,t\right)\psi_{i_1}^\ast \left(\sum_n \tilde{u}_{\alpha,n} \phi_n^\prime,t\right),
\end{eqnarray}
\begin{eqnarray}
\rho_\mathrm{Ir}^{i_2,i_2}(t)&=&\frac{\sqrt{N}}{2}|i_2\rangle \langle i_2| \otimes\int d\phi_{i_1} d\phi_{i_1}^{\prime}d\phi_{i_2} d\phi_{i_2}^{\prime}|\phi_{i_1}\rangle \langle \phi_{i_1}^{\prime}| \otimes |\phi_{i_2}\rangle \langle \phi_{i_2}^{\prime}| \cr
&&\times \prod_{n\neq i_1,i_2}\int d\phi_n \prod_{\alpha=1}^{N-1} \psi_{i_2} \left(\sum_n \tilde{u}_{\alpha,n} \phi_n,t\right)\psi_{i_2}^\ast \left(\sum_n \tilde{u}_{\alpha,n} \phi_n^\prime,t\right),
\end{eqnarray}
\begin{equation}
\rho_\mathrm{Ir}^{i_2,i_1}(t)=\rho_\mathrm{Ir}^{i_1,i_2 \ast}(t).
\end{equation}
If we trace out the ion site state, we obtain the reduced state of the $(i_1, i_2)$ ruler dipole state:
\begin{eqnarray}
\rho_\mathrm{r}(t)&=&\frac{\sqrt{N}}{2}\int d\phi_{i_1} d\phi_{i_1}^{\prime}d\phi_{i_2} d\phi_{i_2}^{\prime}|\phi_{i_1}\rangle \langle \phi_{i_1}^{\prime}| \otimes |\phi_{i_2}\rangle \langle \phi_{i_2}^{\prime}|\cr
&&\times \prod_{n\neq i_1,i_2}\int d\phi_n\prod_{\alpha=1}^{N-1} \left[\psi_{i_1} \left(\sum_n \tilde{u}_{\alpha,n} \phi_n,t\right)\psi_{i_1}^\ast \left(\sum_n \tilde{u}_{\alpha,n} \phi_n^\prime,t\right)\right.\cr
&&\left.+\psi_{i_2} \left(\sum_n \tilde{u}_{\alpha,n} \phi_n,t\right)\psi_{i_2}^\ast \left(\sum_n \tilde{u}_{\alpha,n} \phi_n^\prime,t\right)\right].\label{reducedeq}
\end{eqnarray}

The joint probability density of finding the $(i_1,i_2)$ ruler dipoles at locations $(\phi_{i_1},\phi_{i_2})$ is
\begin{eqnarray}
p(\phi_{i_1},\phi_{i_2},t)&=&\frac{\sqrt{N}}{2}\prod_{n\neq i_1,i_2}\int d\phi_n \prod_{\alpha=1}^{N-1} \left[\left|\psi_{i_1} \left(\sum_n \tilde{u}_{\alpha,n} \phi_n,t\right)\right|^2\right.\cr
&&\left.+ \left| \psi_{i_2} \left(\sum_n \tilde{u}_{\alpha,n} \phi_n,t\right)\right|^2 \right],
\label{jointdensity}
\end{eqnarray}
while the probability density of finding the $j$th ruler dipole at location $\phi_j$ has the folllowing simpler form:
\begin{eqnarray}
p(\phi_j,t)&=&\frac{\sqrt{N}}{2}\prod_{n\neq j}\int d\phi_n \prod_{\alpha=1}^{N-1} \left[\left|\psi_{i_1} \left(\sum_n \tilde{u}_{\alpha,n} \phi_n,t\right)\right|^2\right.\cr
&&\left.+ \left| \psi_{i_2} \left(\sum_n \tilde{u}_{\alpha,n} \phi_n,t\right)\right|^2 \right].
\end{eqnarray}
%\bibliographystyle{quantum}
%\bibliography{biblioRuler}
\end{document}